\documentclass[12pt]{elsarticle}

\journal{Nuclear Inst. and Meth. in Phys. Res., A, NIMA-D-14-00850}

\usepackage{a4wide}
\usepackage{calc}
\usepackage{fancyhdr}
\usepackage{graphicx}
\usepackage[english]{babel}
\usepackage[bookmarks]{hyperref}
\usepackage{multirow}
\usepackage{lmodern}
\usepackage[usenames, dvipsnames]{color}
\usepackage{tabularx}
\usepackage{floatrow}
\usepackage{rotating}
\usepackage{longtable}
\usepackage{booktabs}
\usepackage{amssymb}
\usepackage{amsmath}
\usepackage{enumitem}
\usepackage{gensymb}
\usepackage{amsfonts}
\usepackage{pdflscape}
\usepackage{bm}
\usepackage[mathscr]{euscript}

\usepackage[%
    format=hang,    
    format=plain,
    margin=0pt,
    width=0.8\textwidth,
]{caption}
\usepackage[list=true,listformat=simple]{subcaption}
\usepackage{url}
\usepackage{hyperref}
\usepackage{epstopdf}

\begin{document}


\begin{frontmatter}



\title{Characterisation of an n-type segmented BEGe detector}

\author[a]{I.~Abt}
\author[a]{A.~Caldwell}
\author[a,b]{B.~Donmez}
\author[c]{C.~Etrillard} 
\author[a]{C.~Gooch} 
\author[a]{L.~Hauertmann} 
\author[c]{M.O.~Lampert}
\author[a,d]{H.~Liao}
\author[a]{X.~Liu}
\author[e]{H.~Ma}
\author[a]{B.~Majorovits}
\author[a]{O.~Schulz}
\author[a]{M.~Schuster\corref{cor1}}
\ead{schuster@mpp.mpg.de} 
\cortext[cor1]{Corresponding Author}
\address[a]{Max-Planck-Institut f\"ur Physik, M\"unchen, Germany}
\address[b]{now at University of Antalya, Turkey}
\address[c]{Mirion France, Lingolsheim, France}
\address[d]{now at Kansas State University, Manhattan, USA}
\address[e]{Tsinghua University, Beijing, China}

\begin{abstract}
A four-fold segmented n--type point-contact ``Broad Energy'' high-purity
germanium detector, SegBEGe,  
has been characterised 
at the Max--Planck--Institut f\"ur Physik in Munich.
The main characteristics of the detector are described and
first measurements concerning the detector properties are presented.
The possibility to use mirror pulses to determine source positions
is discussed as well as charge losses observed close to the core contact.
\end{abstract}

\end{frontmatter}

\section{Introduction}
\label{section:introduction}
Germanium detectors are used in a wide 
variety of  
scientific applications~\cite{Vetter2007:RecentDevelopmentsFabrication}, 
in fields like medicine, homeland security and applied and fundamental
physics~\cite{Akkoyun2011:AGATAAdvancedGamma,Abgrall2014:MajoranaDemonstratorNeutrinoless,Ackermann2013:Gerdaexperimentsearch,Aalseth2013:CoGeNTSearchLow}. 
``Broad Energy'' Germanium (BEGe) detectors have become increasingly important
in searches for neutrinoless double beta
decay~\cite{Agostini2015:Productioncharacterizationoperation,Mertens2015:Majoranaexperience}.
The main challenge for these searches is the reduction of background.
This requires as perfect an understanding of the detector properties as
possible.

The segmented n-type BEGe detector presented here was designed 
in order to study
the properties of BEGe detectors in general.
One interesting subject is the influence of the crystal axes on the 
trajectories of the charge carriers 
and thus the pulse shapes. This is linked to the mobility tensors of holes 
and electrons.
The mirror pulses, as observed in segments which do not collect charge,
are an important source of information about the drifting charges inside
the detector and provide spatial information.
The response of  BEGe detectors to interactions close to the 
core contact area,
where charge collection inefficiencies are expected and not well understood,
is another important issue. 
Results are presented for scans of the mantle and the end-plates 
of the subject detector with a $^{133}$Barium source.

\section{The Detector}
\label{section:detector} 

\begin{figure}[ht]
\centering
 \begin{subfigure}{.50\textwidth}
  \centering
  \includegraphics[width=1.0\textwidth]{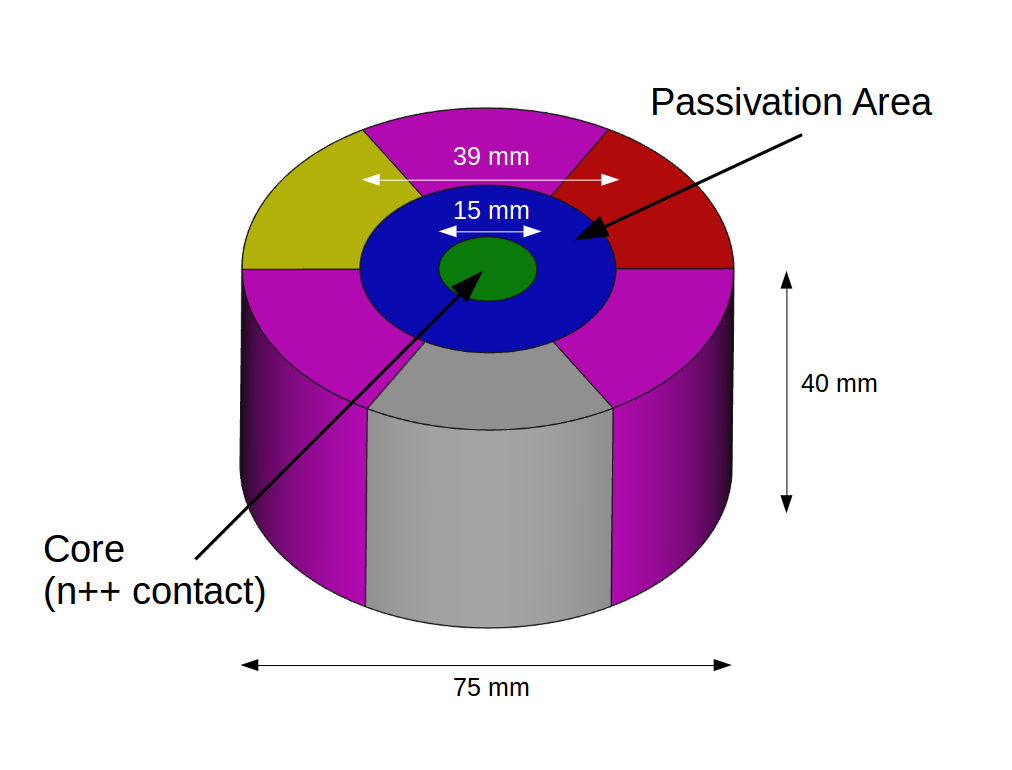}
  \end{subfigure}%
 \begin{subfigure}{.50\textwidth}
  \centering
  \includegraphics[width=1.0\textwidth]{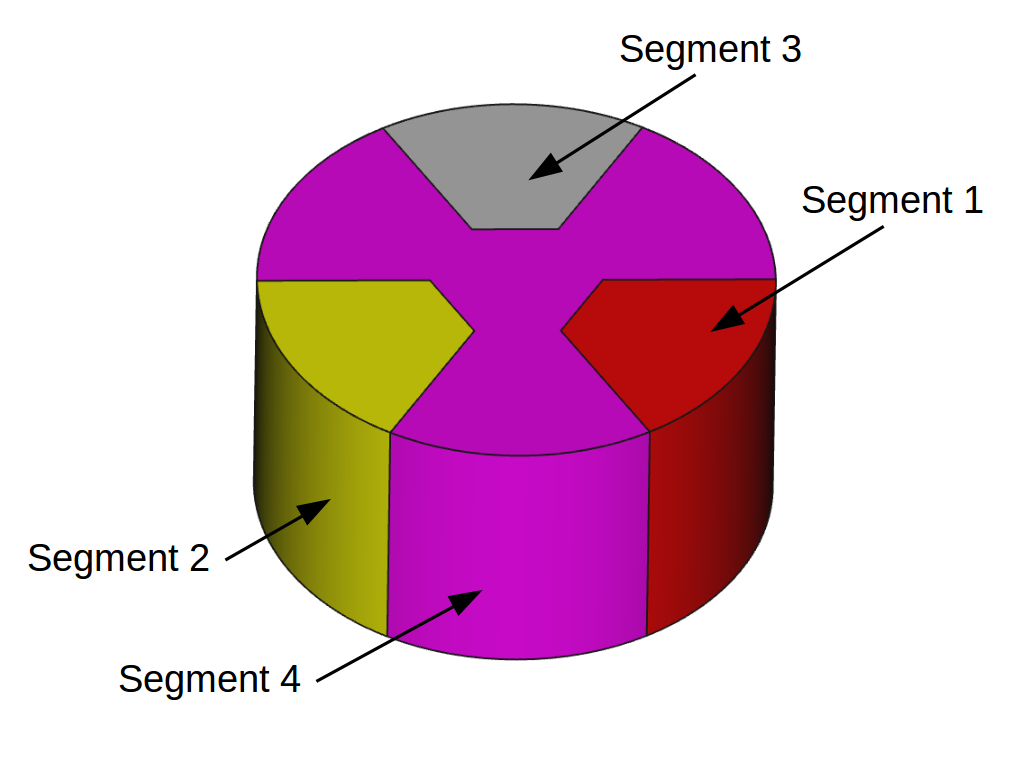}
  \end{subfigure}%
  \vskip -4.3cm \hskip -3.1cm $\phi=0$
  \vskip 4.2cm
\caption {Schematics of the SegBEGe detector seen 
          from the top and (left) 
          from the bottom (right).}
\label{fig:det}
\end{figure}

The ``SegBEGe'' detector is an n-type high-purity 
Broad Energy Germanium (BEGe) detector, segmented
as depicted in Fig.~\ref{fig:det}. 
It has a diameter of 75\,mm and a height of 40\,mm.
Its specifications as provided by the producer,
Mirion France, formerly Canberra France,
are listed in Table~\ref{tab:det:can}.
The side with the n$^{++}$ HV contact, i.e.\ the core contact,
is called the top of the detector.
The core contact has a diameter of 15\,mm and is surrounded by a passivated
ring with an outer diameter of 39\,mm. 
The detector is 
four--fold segmented in $\phi$ with three individual 
60-degree segments, i=1,2,3,
and one segment, segment~4, combining the three other regions in $\phi$.
Segment~4 is closed on the bottom end-plate, see Fig.~\ref{fig:det}.
The segmentation is created through a three-dimensional implantation process.
The center of the bottom plate is the origin of a cylindrical coordinate
system with the $z$-axis pointing towards the core contact.
The left edge of segment~1, looking from the top, defines
$\phi=0$.

\begin{table}[htbp]
\begin{center}
\begin{tabular}{|l|r|}
\hline
Parameter & Value \\ \hline
crystal Diameter& 7.5\,cm\,\,\\
crystal Height  & 4.0\,cm\,\,\\
active Volume   & 177\,cm$^3$\\
\hline
bulk  & n-type\\
effective impurities top & $1.3 ~\,\times 10^{10} /$\,cm$^3$ \\
effective impurities bottom & $0.95 \times 10^{10} /$\,cm$^3$\\
operating voltage & 4500\,V\\
\hline
FWHM at 122 keV & \\
core       & 1.0\,keV\\
segment~1/~2/~3  & 1.9/~2.0/~2.1\,keV\\
segment~4  & 3.7\,keV\\
\hline
FWHM at 1332 keV &  \\
core  & 4.4\,keV\\
segment~1/~2/~3  & 3.7/~3.8/~4.2\,keV\\
segment~4  & 5.5\,keV\\
\hline
\end{tabular}
\caption{Specifications of the SegBEGe detector as provided 
by the manufacturer.}
\label{tab:det:can}
\end{center}
\end{table}

\begin{figure}[ht]
\centering
 \begin{subfigure}{.5\textwidth}
  \centering
  \includegraphics[width=1.0\textwidth]
           {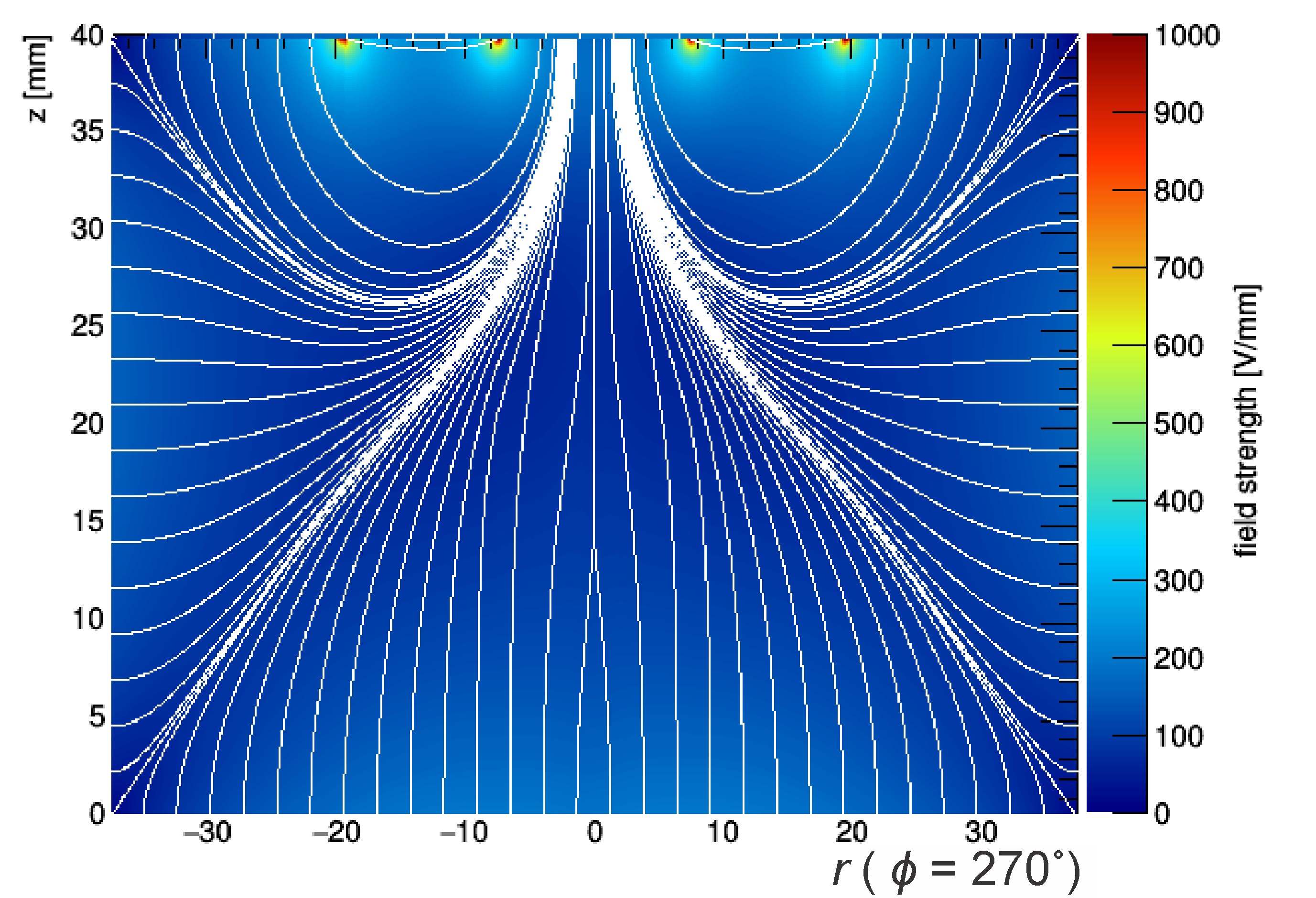}
  \end{subfigure}%
 \begin{subfigure}{.5\textwidth}
  \centering
  \includegraphics[width=1.0\textwidth]
            {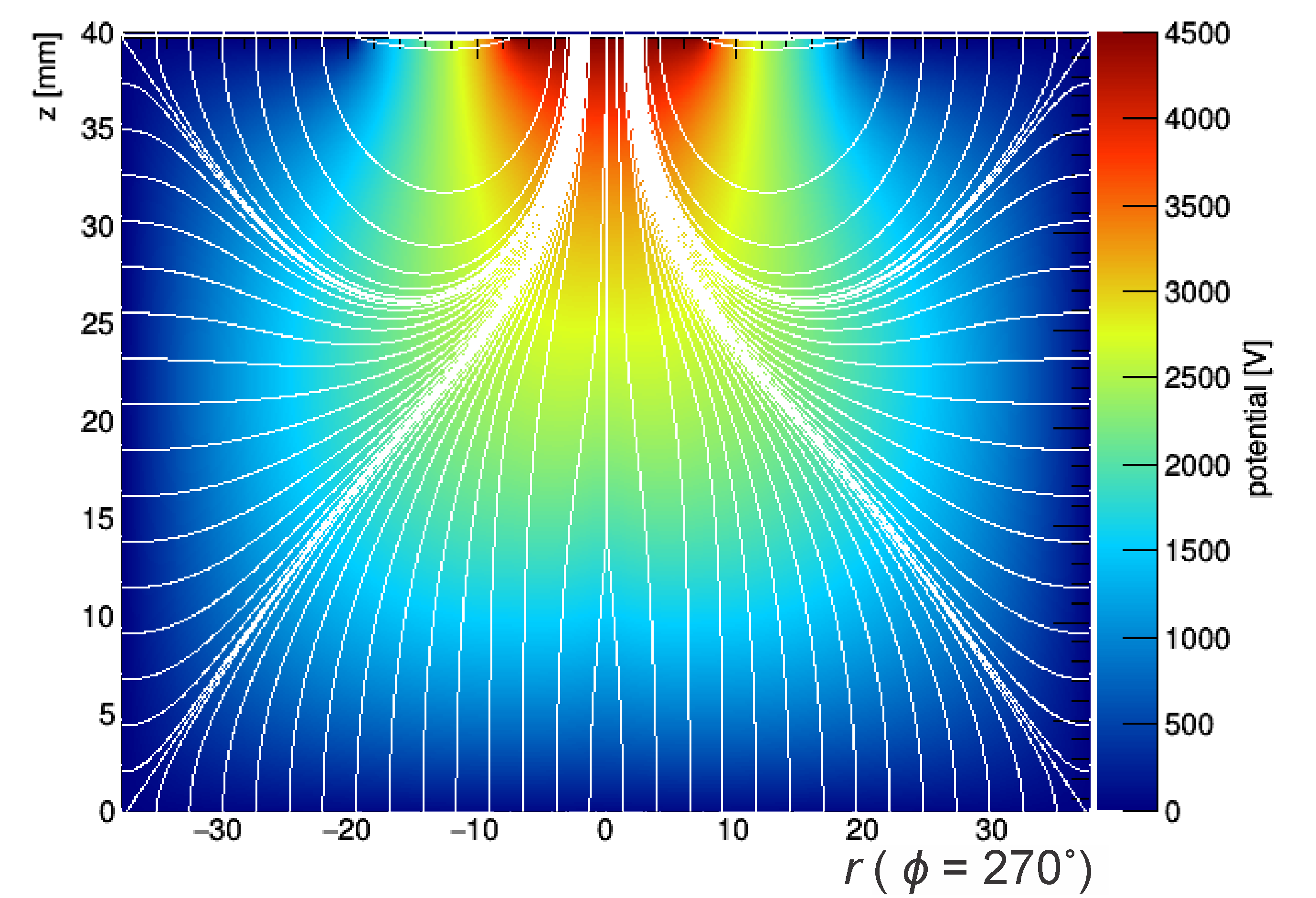}
  \end{subfigure}%
\vspace{0.5cm}
\caption {Electric field strength (left) and potential (right) for 
          the  $r-z$ cut through the detector at $\phi = 270\degree$.
          Also shown are the field lines.}
\label{fig:field}
\end{figure}

The electric field inside the detector is very similar to the field 
of an unsegmented detector. It was calculated using an upgrade
to the program package described in an earlier 
publication~\cite{Abt2010:Pulseshapesimulation}.
The main improvements are the implementation of an adaptive grid and
realistic segment boundaries.
The field distortions on the mantle
close to the surface around the narrow segment boundaries were found
to be very shallow and insignificant for all gamma scans.
The effect of the shape of the segmentation on the bottom plate  
on the field lines is visible in Fig.~\ref{fig:field} which shows
the electrical field strength and the potential as well as 
the field lines for the  $r-z$ cut through the detector 
at $ \phi = 270\degree$. 
The slight asymmetry of the field lines seen around $r=0$  
is caused by the influence of the inner boundary of segment~3.
The ``positive radii'' in Fig.~\ref{fig:field}
indicate the cut through the middle of segment~3, the ``negative radii''
indicate the cut through segment~4, which also covers the centre of the
bottom plate, see Fig.~\ref{fig:det}.
Figure~\ref{fig:field} is based on calculations where the width of 
the floating segment boundaries was 
assumed to be 1\,mm, which is not the precise width but serves 
to demonstrate the influence of the segment boundaries.

The potential 
close to the core contact of such a detector is high.
It drops rapidly, creating a strong field close to the core contact while
the field close to the mantle of the detector is weak.
This causes the large differences in drift speed for different
regions typical for this type of detector.

The field strength at the edges of the passivated ring around
the HV contact is not expected to be as high as indicated by the
calculation. The calculation is based entirely on the boundary conditions
for the potential which are 0\,V for the segments, 4500\,V for the core
contact and ``floating'' for the passivated ring.
Neither the depth of the Lithium drifted core contact nor the 
Lithium diffusion
at the edge were implemented. Similarly, no depth or diffusion was
implemented for the boron implants of the segments.
Especially, the diffusion is expected to reduce the spikes in the
field strength.

\section{The Experimental Setup}
\label{section:setup} 

\begin{figure}[ht]
\vspace{-0.3cm} 
\centerline{
\includegraphics[width=0.55\textwidth,angle=0]{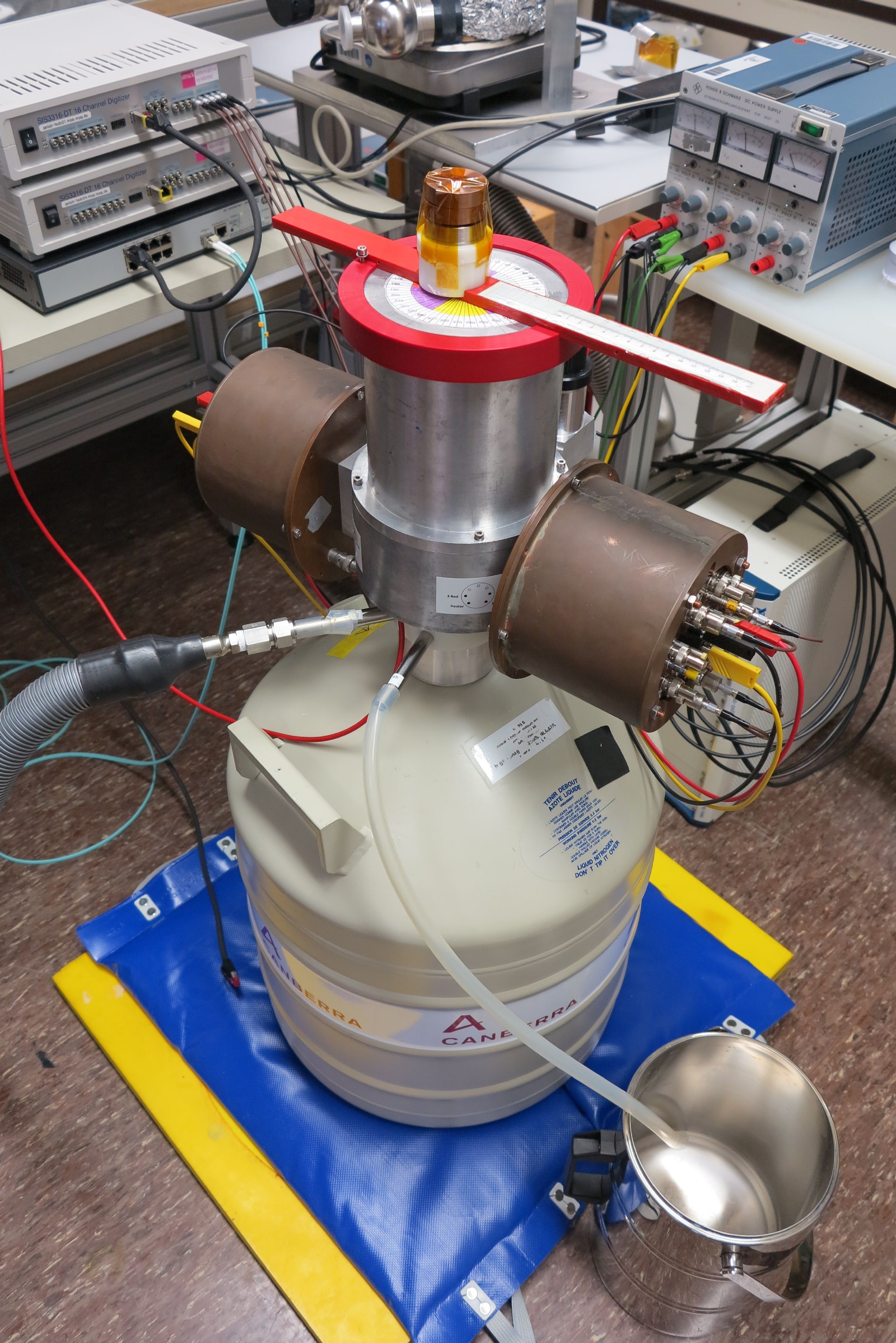}}
\caption {The experimental setup during a scan of the top of the
          detector during run~B, figure taken from~\cite{MTh:Schuster2017}.
          The cryostat is mounted on top of a liquid nitrogen
          dewar resting on a
          sandbag to reduce vibrations. Copper ``ears'' house the
          readout electronics. The source was positioned manually
          on a grid. 
}
\label{fig:K1}
\end{figure}

The detector was mounted in
a conventional aluminium vacuum-cryostat called K1
for the characterisation measurements presented here. 
This cryostat was used previously to study the
performance of the first 18-fold segmented 
true coaxial detector~\cite{Abt2007:Characterizationfirsttrue}.
In K1, the detector is cooled through a 
copper finger submerged in a conventional liquid nitrogen dewar. 
The temperature at the top of the cooling finger
was monitored using a PT100 inside the vacuum cap. 
Between daily refilling, the temperature was stable between
102\,K and 106\,K.
Any influence due to changes in temperature was not corrected for
in the studies presented here.

\begin{figure}[tbp]
\centerline{
\includegraphics[width=0.7\textwidth]{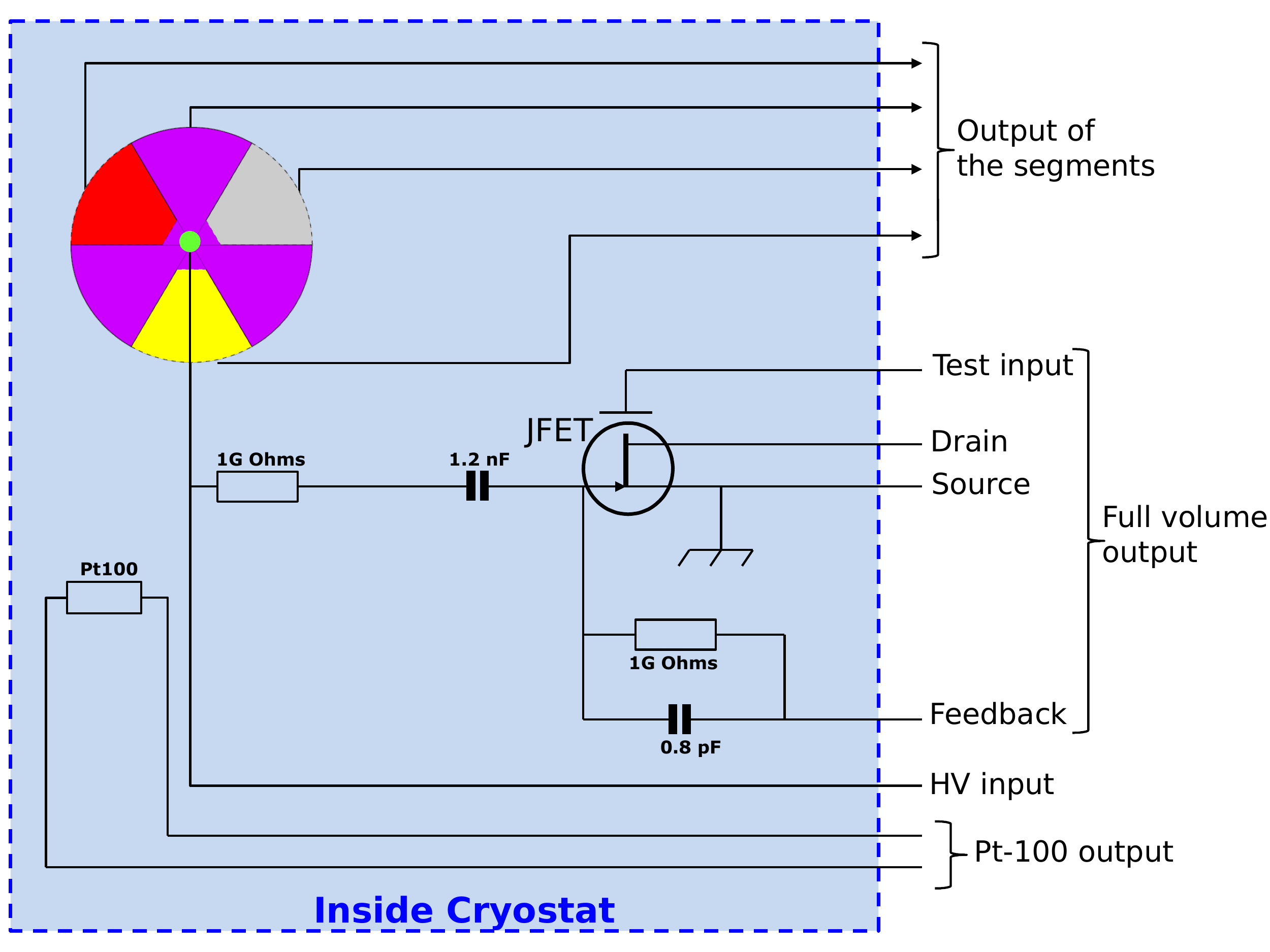}}
\vspace{0.5cm}
\caption {Schematic of the detector readout, 
          taken from~\cite{MTh:Schuster2017}.
          The segment signals were processed in one ear and the
          core (full volume) signal in the other ear.}
\label{fig:readout}
\end{figure}

The setup is depicted in Fig.~\ref{fig:K1}.
The signals were amplified by PSC~823~pre-amplifiers produced
by Mirion France, which were housed in the copper
``ears''  visible in Fig.~\ref{fig:K1}. 
The DC-coupled room-temperature FETs for the signals of the
segments were sitting on the boards of their four pre-amplifiers
in one of the ears.
The cold FET for the AC-coupled core signal was located inside the 
detector cap and was thermally coupled to the cold finger. The rest
of the core pre-amplification stage was located in the other ear.  
The schematic of the readout is given in Fig.~\ref{fig:readout}.

The detector was 
first mounted upside down, i.e.\ with the core contact down;
this period is called run~A. 
For the following run~B, it was mounted with the core contact up.
The data acquisition systems were different for runs~A and~B.
For run~A, a PIXIE-4 system~\cite{Pixie4} with a 75\,MHz sampling
frequency and a 13.7\,$\rm{\mu}$s trace length was used. 
This system used a trapezoidal filter for threshold triggering.
It also provided internal pile-up suppression. 
For run~B, a Struck 16-channel SIS3316-250-14 module~\cite{def:2014} 
with a sampling frequency of 250\,MHz 
and a trace length of 20\,$\mu$s was used. 
This system used a trapezoidal filter with additional
constant fraction time-positioning for threshold triggering.
It did not provide any online suppression of
saturated or pile-up events.

\section{Data Taking}

The detector was first commissioned  in the fall of 2014.
Run~A lasted until summer 2015.
Run~B, with the detector upright, lasted from 
March to April 2016.

In both run periods, data were taken with an uncollimated  $^{60}$Cobalt
and an uncollimated $^{228}$Thorium source to illuminate the detector bulk
and study resolutions.
The detector scans were performed with a $^{133}$Barium source.
Side-scans were performed in both runs. In run~A (B), also
the bottom (top) of the detector was scanned.
The data sets which were used for this paper
are listed in Table~\ref{tab:run:A+B}.

The gammas from the $^{133}$Barium source were collimated with a 
50\,mm long tungsten collimator with a diameter of 35\,mm 
and a 1\,mm radius collimation hole.
A purely geometrical calculation shows that the 
beam spots on the detector surface
had radii of 2.9\,mm on the side and 2.6\,mm on the end-plates.

\begin{table}[h]
\renewcommand\arraystretch{1.1}
\begin{center}
\begin{tabular}{|l|l|l|c|l|l|c|}
\hline
Label & Type & Source & Beam Spot & $z$ & $r$ & $\phi$\\ \hline
{\bf Co-B}  & bulk & $^{60}$Co   & n/a & on top of& $r$=0 & n/a\\
{\bf Th-B}  & bulk & $^{228}$Th  & n/a & cryostat & $r$=0 & n/a\\
\hline
\hline
{\bf Bsc}  & bottom-scan & $^{133}$Ba & 2.6\,mm &  
                                  $z$=0\,mm & $r$=25.0\,mm & 0 -- 360 $^o$\\
\hline
{\bf Ssc-A} & side-scan & $^{133}$Ba & 2.9\,mm &  
                                  $z$=20\,mm & $r$=37.5\,mm & 0 -- 360 $^o$\\
\hline
{\bf Ssc-B} & side-scan & $^{133}$Ba & 2.9\,mm &  
                                  $z$=20\,mm & $r$=37.5\,mm & 0 -- 360 $^o$\\
\hline
{\bf Tsc}  & top-scans & $^{133}$Ba & 2.6\,mm &  
                                  $z$=40\,mm & $r$=30,18, & 0 -- 360 $^o$\\
           &          &           &         &  
                                             & ~~~13,9,6\,mm   &          \\
\hline
\hline
\end{tabular}
\caption{Data sets used for this paper.
The values listed under beam spot are the beam spot radii
on the detector surface.
The $z$, $r$ and $\phi$ values listed are in detector coordinates.
The $z$ and $r$ values were adjusted to $\pm$\,0.5\,mm, $\phi$ was controlled
to $\pm$\,1\,degree. 
}
\label{tab:run:A+B}
\end{center}
\end{table}

\section{Data Processing}

The online energy determination provided by the PIXIE system in 
run~A was not used for the results presented here. 
However, the PIXIE system suppressed pile-up and saturated events
online while the Struck system was not programmed to suppress such events.
For run~B, events from pile-up and
saturation were suppressed by evaluating the slopes of the baseline
and the decay-corrected signal plateau.
A negative baseline slope indicates pile-up and a positive plateau slope
indicates saturation. During Barium scans, a total of about 5\,\% of the
events were rejected. These  were dominated by events
with a saturated amplifier. Actual pile-up was at the level of about 1\,\%. 

The rest of the offline data processing was identical for both 
runs.
The recorded raw pulses were baseline subtracted and corrected for
the pre-amplifier specific decay of the 
pulses~\cite{MTh:Schuster2017,MTh:Hauertmann2017}.
Signal amplitudes were derived using a fixed-size window filter, 
where the position of the window was determined by the trigger.
For run~A~(B), 
a baseline window of 4.48\,$\mu$s (8.0\,$\mu$) and
an amplitude window of 6.63\,$\mu$s (9.0\,$\mu$) were chosen.
This filter introduces the least bias with respect to different pulse
shapes.
The baseline and amplitude windows were separated far enough to ensure
that the actual rise of the pulse started after the baseline and ended
before the amplitude window.

Cross-talk effects between the core and the segments 
as well as between segments
were treated  in an automated calibration procedure
using single-segment events~\cite{MTh:Schuster2017,MTh:Hauertmann2017}.
A calibration was performed for each data set individually.
The cross-talk correction was performed under the assumption 
that the cross-talk from all segments to the core was identical.
This assumption affected the core energy scale 
for single-segment events in different segments by less
than 0.1\,\%~\cite{MTh:Schuster2017,MTh:Hauertmann2017}.
In effect, this insignificantly worsens the core energy resolution
and does not affect any aspect of the analysis.
The cross-talk from one particular segment to another segment is always
measured together with the cross-talk from the core to this segment.
Segment~4 events resulted in a cross-talk of $1\sim 2$\,\% into 
the small segments~1,2 and~3 while events in these small segments caused a
cross-talk of about 0.4\,\% into the large segment~4.

\section{Overall Detector Performance}
  
\begin{figure}[h]
\vskip -0.5cm
\centerline{
\includegraphics[width=0.6\textwidth]
        {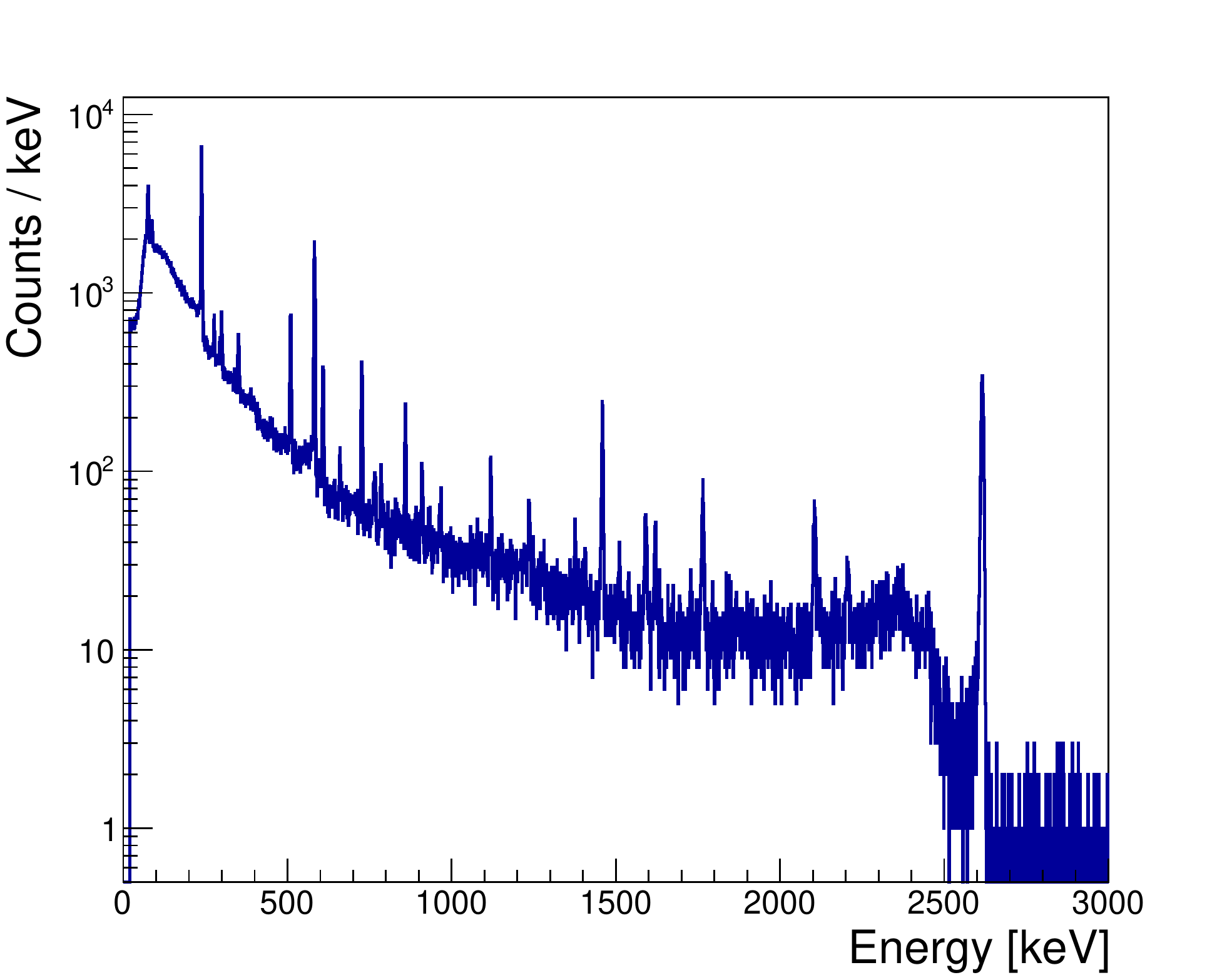}}
\caption{Core spectrum from a 1-hour {\bf Th-B} data set. 
         }
\label{fig:co:spectrum}
\end{figure}

\begin{table}[h]
\begin{center}
\begin{tabular}{|c|c||c|c|c|c|c|}
\cline{3-7}
\multicolumn{2}{c}{}&\multicolumn{5}{|c|}{\textbf{FWHM~[keV]}}\\
\hline
\textbf{Source}&\textbf{$\gamma$-line~[keV]}&\textbf{Core}&\textbf{Seg. 1}&\textbf{Seg. 2}&\textbf{Seg. 3}&\textbf{Seg. 4}\\
\hline
\hline
\multirow{2}{*}{$^{133}$Ba}&81&3.29&5.15&4.05&4.07&7.26\\
\cline{2-7}
&356&2.64&4.58&3.47&3.52&6.04\\
\hline
\multirow{2}{*}{$^{60}$Co}&1173&4.57&4.88&4.16&4.34&8.05\\
\cline{2-7}
&1332&4.93&5.00&4.39&4.14&7.84\\
\hline
$^{228}$Th&2614&7.65\\
\cline{1-3}
\end{tabular}
\caption{\label{tab:energy_resolutions}Energy resolutions 
         as absolute FWHMs in keV for the core and all segments
         as observed in run~B.}
\end{center}
\vskip -.3cm
\end{table}

The detector performed as expected in the K1~cryostat. The conditions in this
setup were not perfect with respect to grounding and shielding. 
Thus, the resolutions were a bit worse than listed by the manufacturer.
Results from run~B are listed in Table~\ref{tab:energy_resolutions}.
The lack of energy dependence for the resolutions demonstrates that the results
are dominated by electronic noise. 
However, the detector resolution did not affect any of the results presented
here. Figure~\ref{fig:co:spectrum} shows the core spectrum for a 
1-hour {\bf Th-B} data set. 
The 2614\,keV $^{208}$Tl line is clearly visible as well as 
the $^{212}$Bi lines at 239 and 1620\,keV. 
In addition, the natural background in the laboratory features the 
usual lines from the uranium decay chain as well as a strong 1460\,keV line 
from $^{40}$K.

The double-escape peak from the 2614\,keV Thallium line at 1592\,keV 
and the 1620\,keV Bismuth line were used for a standard pulse-shape
analysis to show that the segmentation did not affect the core pulses.
The Bismuth line is dominated by multi-site events from Compton scattering.
The double-escape peak is dominated by single-site events; in these events 
all the energy is deposited in one small volume. 
The so-called A/E-method uses as a discriminator the ratio of A, 
the maximum of the first derivative of a pulse, i.e.\ the maximum current, 
divided by the total energy of the event.
The method was applied to {\bf Th-B} data~\cite{MTh:Schuster2017}.
For a survival probability of the double-escape peak of 90\,\%,
a reduction of the Bismuth peak of 86\,\% was obtained. That is
compatible with the results obtained for other 
BEGe detectors~\cite{Agostini2015:Productioncharacterizationoperation,PhD:Liao2016}.
\begin{figure}[h]
\centerline{
\includegraphics[width=0.60\textwidth]
  {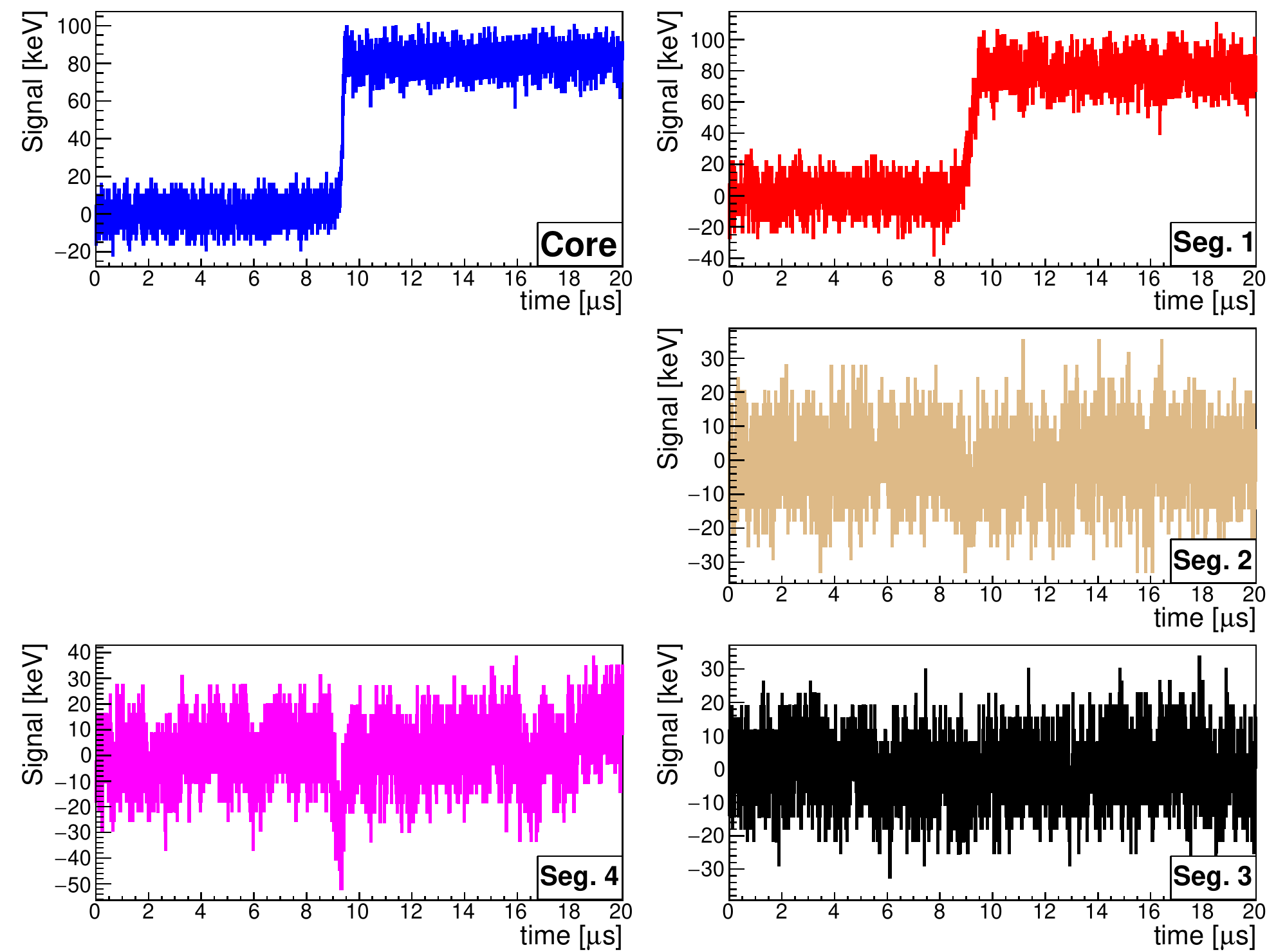}}
\vskip -5.2cm \hskip -5.cm
\includegraphics[width=0.16\textwidth]{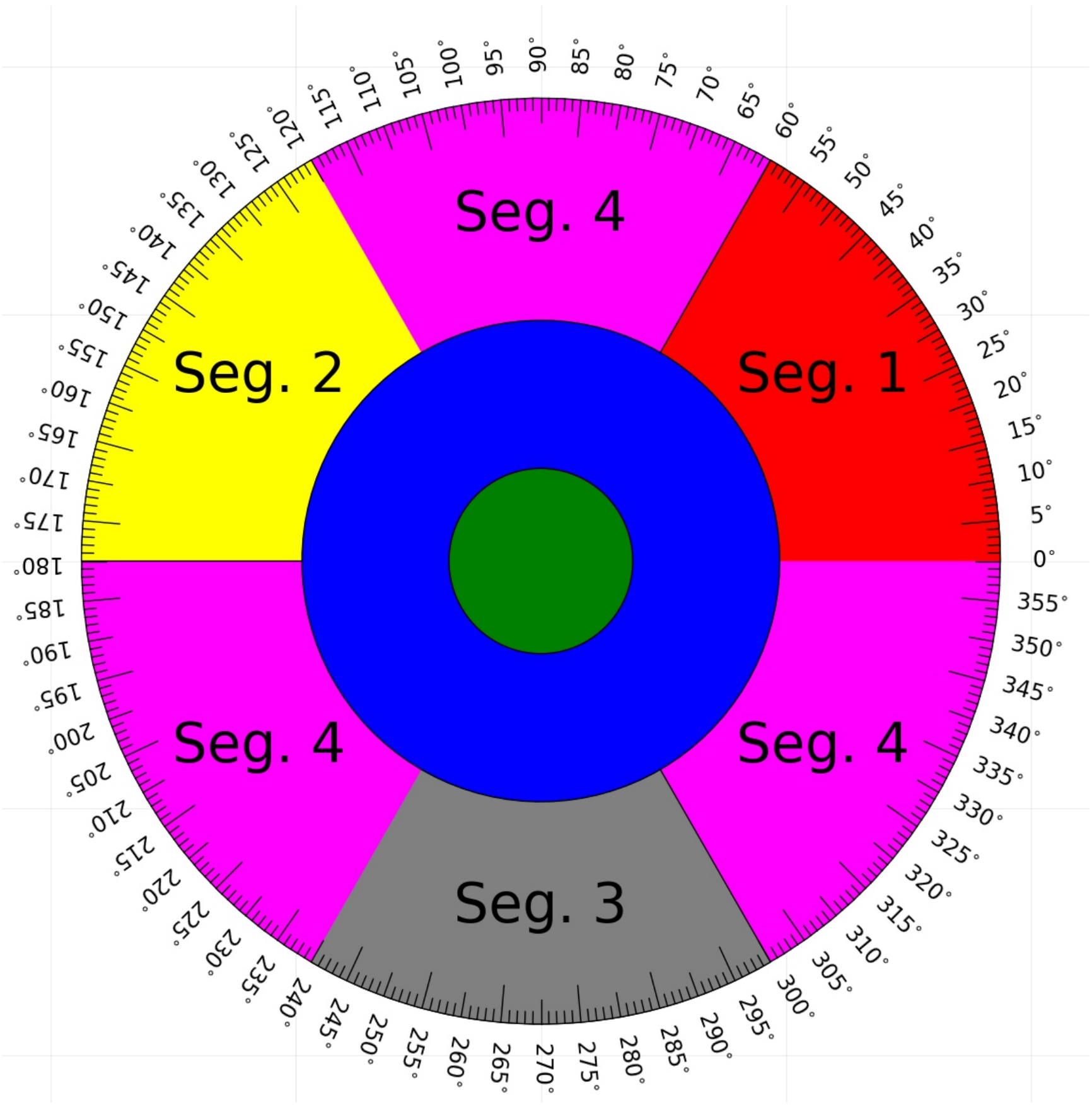}
\vskip +2.6cm 
\caption{Single 81\,keV event from the {\bf Ssc-B} 
         data set for $\phi=45^{\circ}$. The core pulse is shown at the
         top left, segments~1,2,3 are shown from top to bottom on the
         right and segment~4 is shown at the bottom left.
         The inset depicts the detector top with a $\phi$ scale.  
         }
\label{fig:pulse:81keV}

\end{figure}

\section{Super-pulses}

The 81\,keV line from $^{133}$Ba was chosen as the line 
for which scanning results 
are presented. The gammas from this line have a penetration depth of
about 1.8\,mm and thus create events very close to the surface.
As a result, the holes
are collected quickly and the drift is dominated by electrons.
A single event at 81\,keV as recorded in the {\bf Ssc-B} data set 
at $\phi=45^{\circ}$ is shown in Fig.~\ref{fig:pulse:81keV}.

\begin{figure}[h]
\centerline{
\includegraphics[width=0.60\textwidth]
  {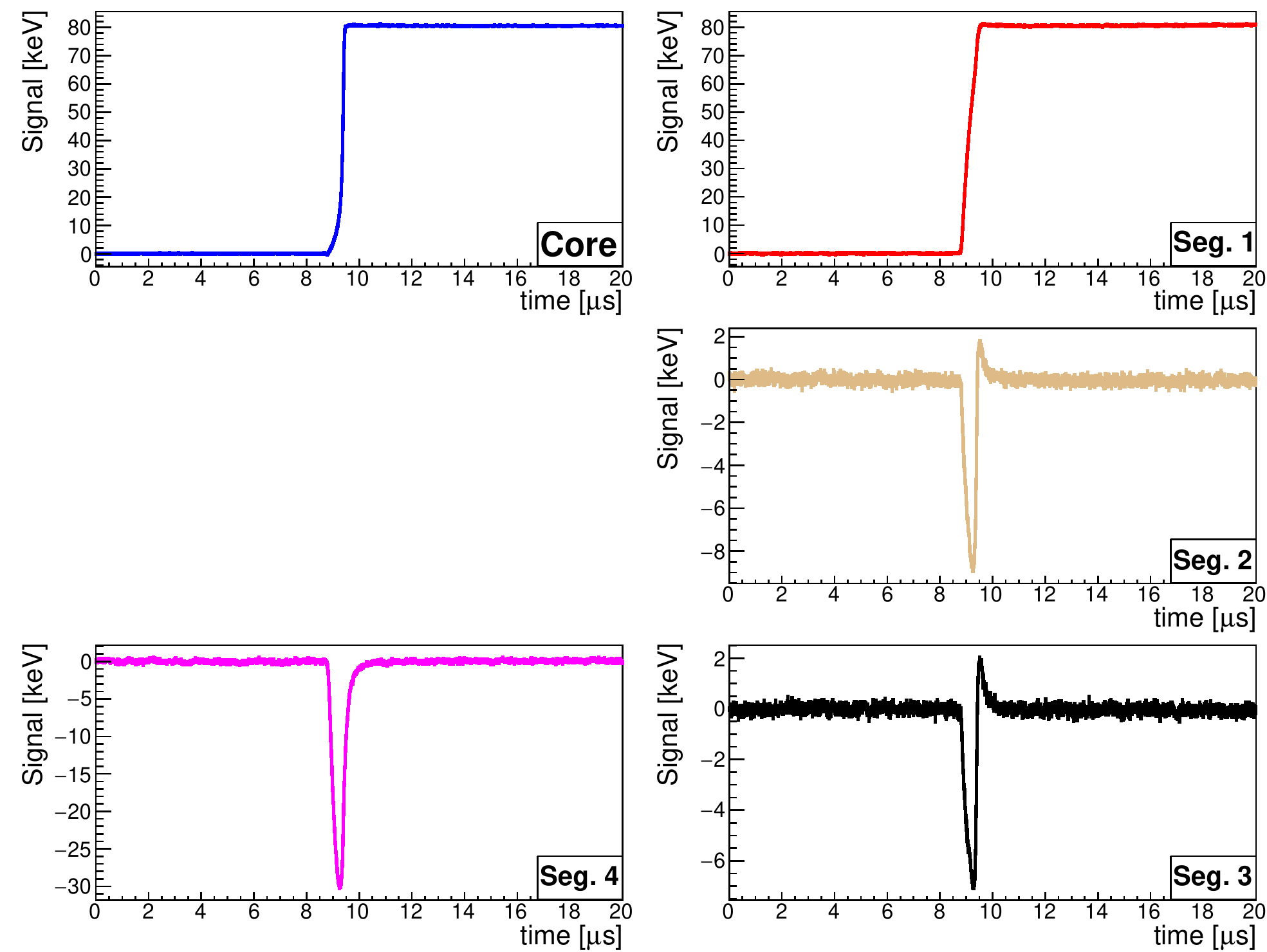}}
\vskip -5.2cm \hskip -5.cm
\includegraphics[width=0.16\textwidth]{Inset-I.pdf}
\vskip +2.60cm 
\caption{Super-pulse for 81\,keV events from the {\bf Ssc-B} 
         data set for $\phi=45^{\circ}$. The core super-pulse is shown at the
         top left, segments~1,2,3 are shown from top to bottom on the
         right and segment~4 is shown at the bottom left.
         Please note the different scales used for the different
         segments. 
         The inset depicts the detector top with a $\phi$ scale.  
         }
\label{fig:superpulse:81keV}
\end{figure}

The event was located on the surface of segment~1.
The largest mirror pulse is expected in segment~4 next to the
collecting segment~1. It is clearly
visible in Fig.~\ref{fig:pulse:81keV}. 
Smaller mirror pulses are expected in
segments~2 and ~3. The noise level is such that they cannot 
be easily identified in individual events.
As the S/N ratio was 10 or more for this line and the pulses are all very
similar due to the low penetration power of the 81\,keV gammas, the pulses
in 10\,keV intervals around the peaks were averaged to form
super-pulses for each position.
Typically, 2000 to 2500 pulses were averaged.
The super-pulse for the location of the event depicted in 
Fig.~\ref{fig:pulse:81keV} is shown in Fig.~\ref{fig:superpulse:81keV}.
The super-pulse also clearly reveals the smaller mirror pulses 
in segments~2 and~3. As the noise gets averaged out
super-pulses are a powerful tool to investigate detector properties.

\section{Segment Boundaries}
\label{sec:bound}

\begin{figure}[h]
\centerline{
\includegraphics[width=0.9\textwidth]
  {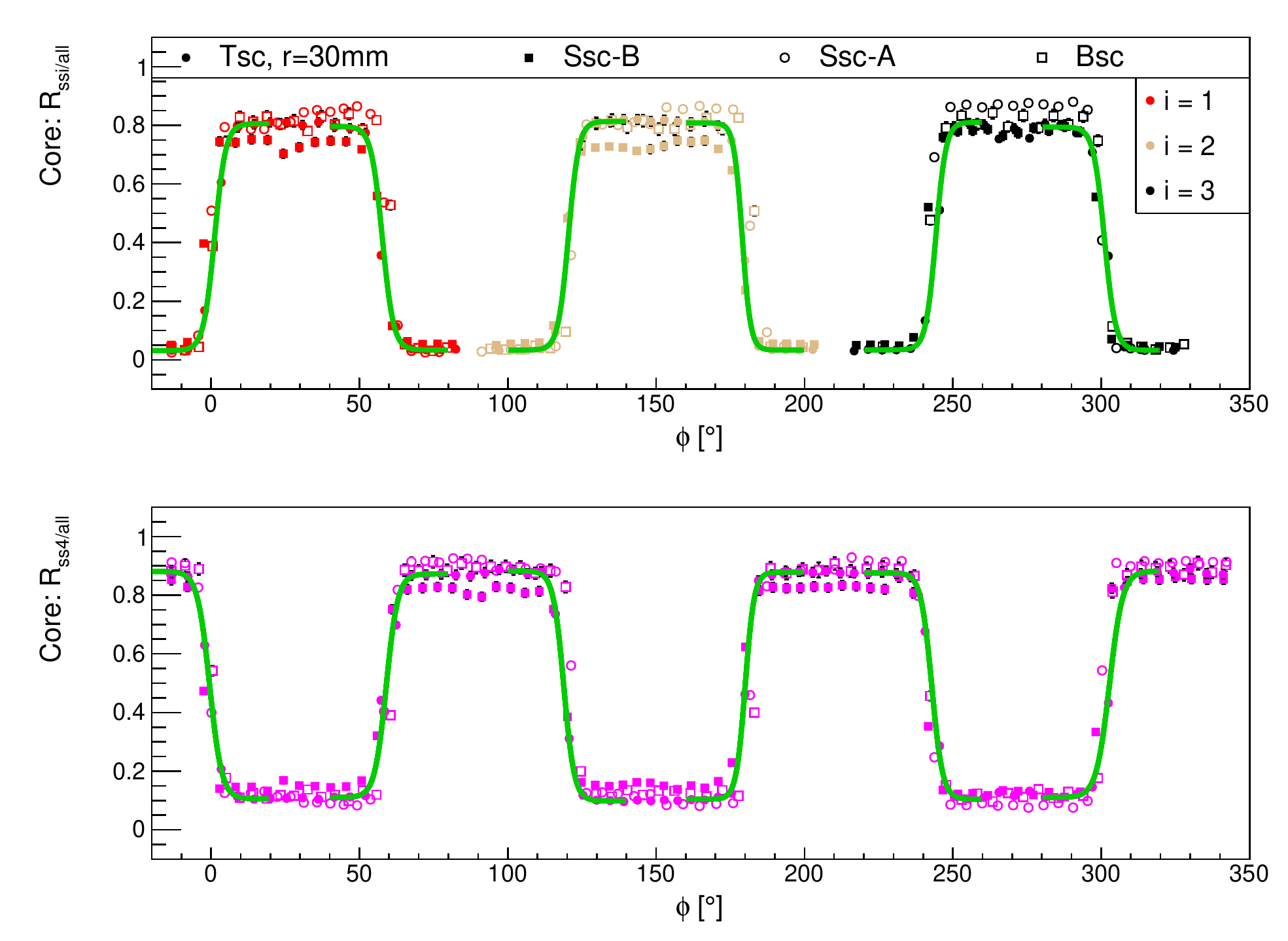}}
\caption{Ratios $R_{ss_i/all}$ for (top) the three individual 
         segments i=1,2,3 and (bottom) the large segment~4.
         Shown are data for run~A (open symbols) and run~B (full symbols). 
         Also shown are fits of the function in Eq.~\ref{eq:sb}
         to the {\bf Tsc} data from run~B.}
\label{fig:seg:bound}
\end{figure}

The segment boundaries were determined using the rate of single-segment 
events in the respective bottom-, side- and top-scans.
The ratios, $R_{ss_i/all}$, of the number of single-segment events
in segment $i$ with $i \in  [1,2,3,4]$ over all single-segment events
were used.
A value close to one is expected if the source is facing the respective 
segment, close to zero, depending on the background level, 
is expected otherwise.
The data as obtained for the 81\,keV line from $^{133}$Ba
in the {\bf Bsc}, {\bf Ssc-A/B}
and {\bf Tsc} 
scans are depicted in Fig.~\ref{fig:seg:bound}. 
Segment~4 has a higher 
background level due to its larger volume. The side-scans were affected
by some parts of the detector holder, 
reducing the event numbers in the middle of
some segments.
Also shown are the results of fits to the {\bf Tsc} data using the function:

\begin{equation}\label{eq:sb}
R_{ss i/all}(\phi) = \frac{H}{2} \cdot 
               \tanh[\Lambda \cdot (\phi - \phi_{i,j})]+\Gamma~~~,
\end{equation}
where the four fitted parameters are
\begin{itemize}
\item
$H$: the maximal variation in $R_{ss i/all}$, 
\item
$\Lambda$: the slope of the variation in $R_{ss i/all}$: $\Lambda > (<)~0$ for
rising (falling) edges,
\item
$\phi_{i,j}$: the boundary between segments $i$ and $j$,
\item
$\Gamma$: the source location independent background.
\end{itemize} 

The segment boundaries were found consistently  in all scans during 
both run periods. The information was mainly used to have
precise location information on the detector.

\section{Crystal Axes}

\begin{figure}[h]
\centerline{
\includegraphics[width=0.8\textwidth]
  {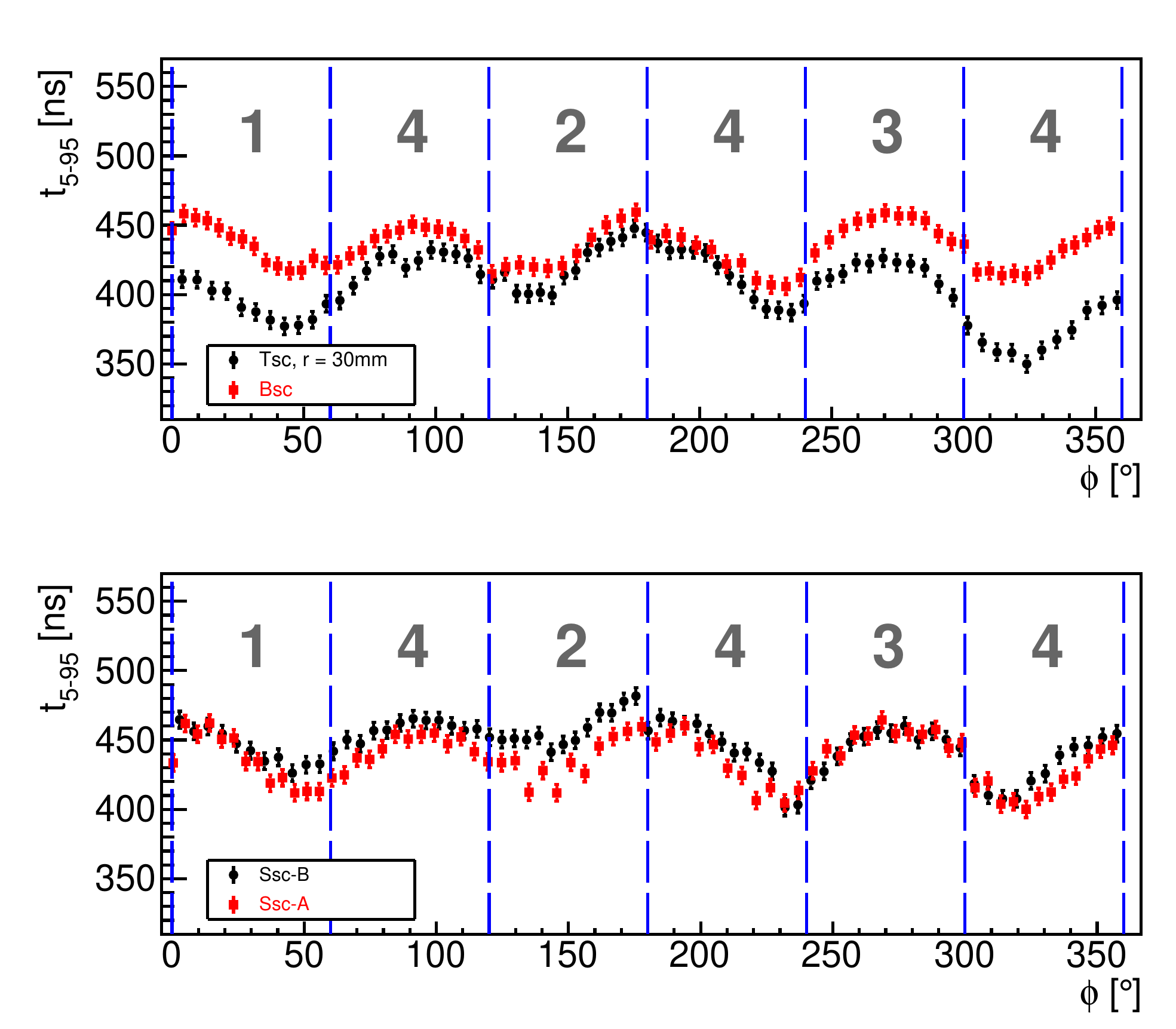}}
\caption{Average 5\,$\%$~to~95\,$\%$ rise-times of 81\,keV super-pulses 
         as a function of the azimuth angle $\phi$ for (bottom) 
         the two side-scans from runs~A and~B and (top) the 
         bottom and top-scans from runs~A and B.
         The error bars represent an uncertainty corresponding 
         to a temperature shift of roughly 2\,K.
         }
\label{fig:crystal-axes}
\end{figure}

The propagation of electrons and holes in the electric field
of the germanium crystal is influenced by the crystal 
axes~\cite{Reik1962:crystalaxis}. 
The charge carriers get deflected and do not follow a simple
radial path.
Thus, the time to collect the charge carriers depends on the angle between
the closest crystal axis and the radial line on which the interaction
takes place.
For a perfect crystal in a perfect setup, a sine function is expected:

\begin{equation}\label{eq:crystal_axis}
t_{5-95}=C + a \cdot \sin \left[ \frac{2\pi}{90} \left( \phi + \phi_{\rm offset}\right)\right]~~~,
\end{equation}

where $t_{5-95}$ is the time a pulse needs to rise from 5\,\% to 95\,\% of 
its amplitude, $C$ is the mean  $t_{5-95}$ and $a$ the amplitude of the variation
of $t_{5-95}$.
The parameter $\phi_{\rm offset}$ is fitted to determine the location of the axes.
 
The data using 81\,keV super-pulses are shown  
in Fig.~\ref{fig:crystal-axes} for scan data from both run~A and~B.
The data from the different periods and scans agree well.
However, the scan data were affected by a small tilt of the detector that
caused a shift of the impact points with respect to the detector coordinates. 
This effect was not corrected for. The top-scan was affected the most.
The smaller radius at which
the bottom-scan was performed limited this effect for the bottom-scan.
The error bars shown in Fig.~\ref{fig:crystal-axes} represent
the uncertainty due to changes in the temperature, which
was not controlled to better than $\pm$\,2\,K.

\begin{figure}[h]
\centerline{
\includegraphics[width=0.85\textwidth]
       {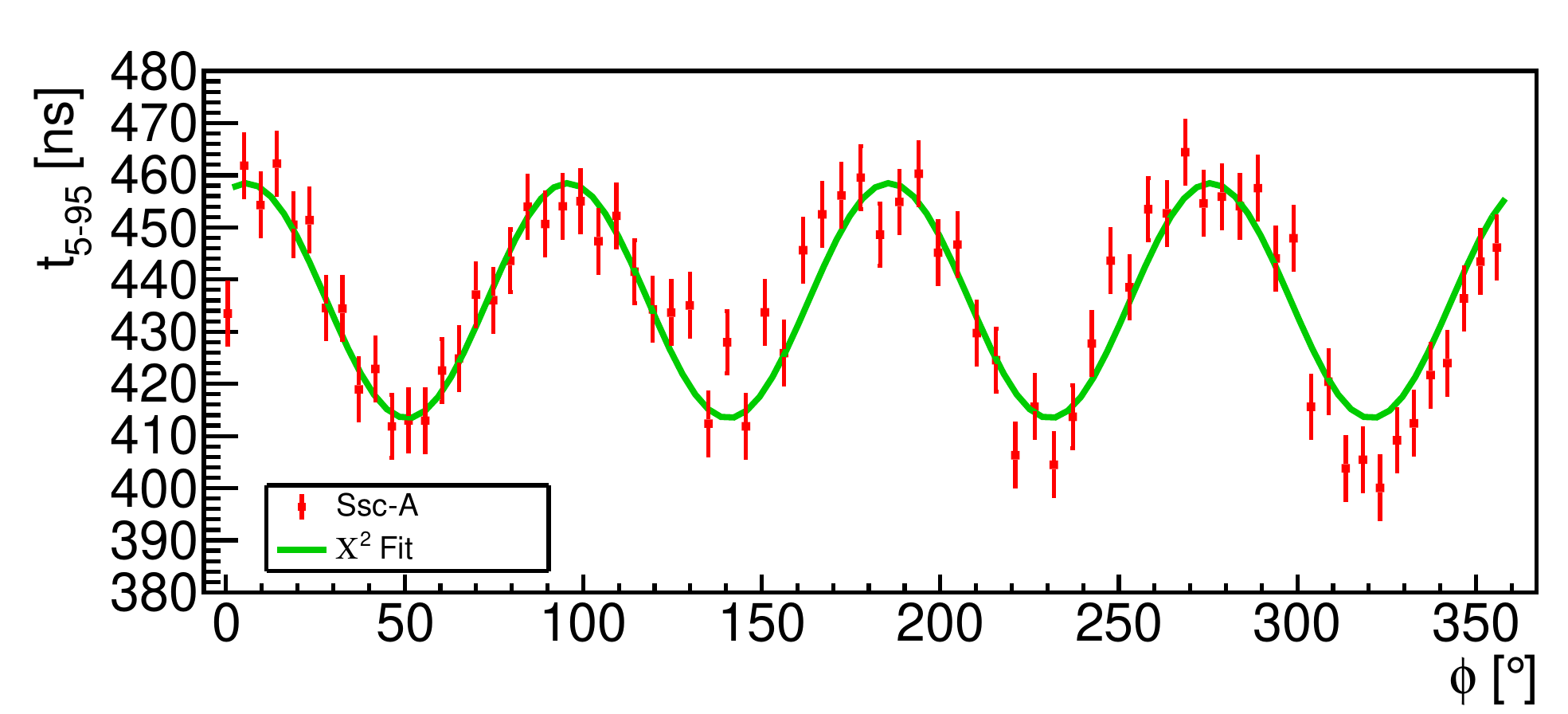}}
\caption{Average 5\,$\%$~to~95\,$\%$ rise-times of 81\,keV super-pulses 
         from Ssc-A together with a fit according to Eq.~\ref{eq:crystal_axis}.
         }
\label{fig:axes:fit}
\end{figure}

The side-scan data from run~A are shown together with a fit
according to Eq.~\ref{eq:crystal_axis} in Fig.~\ref{fig:axes:fit}. 
The sine function describes the data  well. The $\phi$ values
for which the drift-time is maximal indicate the so-called ``slow axes''.
The axes with minimal drift-time are called ``fast axes''.
The difference between drift-times will, in the future, be compared to
simulation results to study the  mobility of electrons.

\section{Position Reconstruction}

The electrons and holes drifting to the electrodes of the collecting segment
create mirror charges in the neighboring segments. These mirror pulses
end at the baseline once the charge carriers are collected at the
electrodes.
This phenomenon can be
understood and deduced from Ramo's theorem.
The amplitudes, MA$_i$ for segments~$i=1,2,3,4$, 
defined as the maximum of the absolute values, 
and shapes of the induced mirror pulses depend
on the location of the interaction point which determines the drift paths
of the charge carriers.
The closer a drifting charge passes a neighbouring segment $k$,
the larger MA$_k$ becomes.

\begin{figure}[h]
\vspace{0.1cm} 
\centerline{
\includegraphics[width=0.8\textwidth]
   {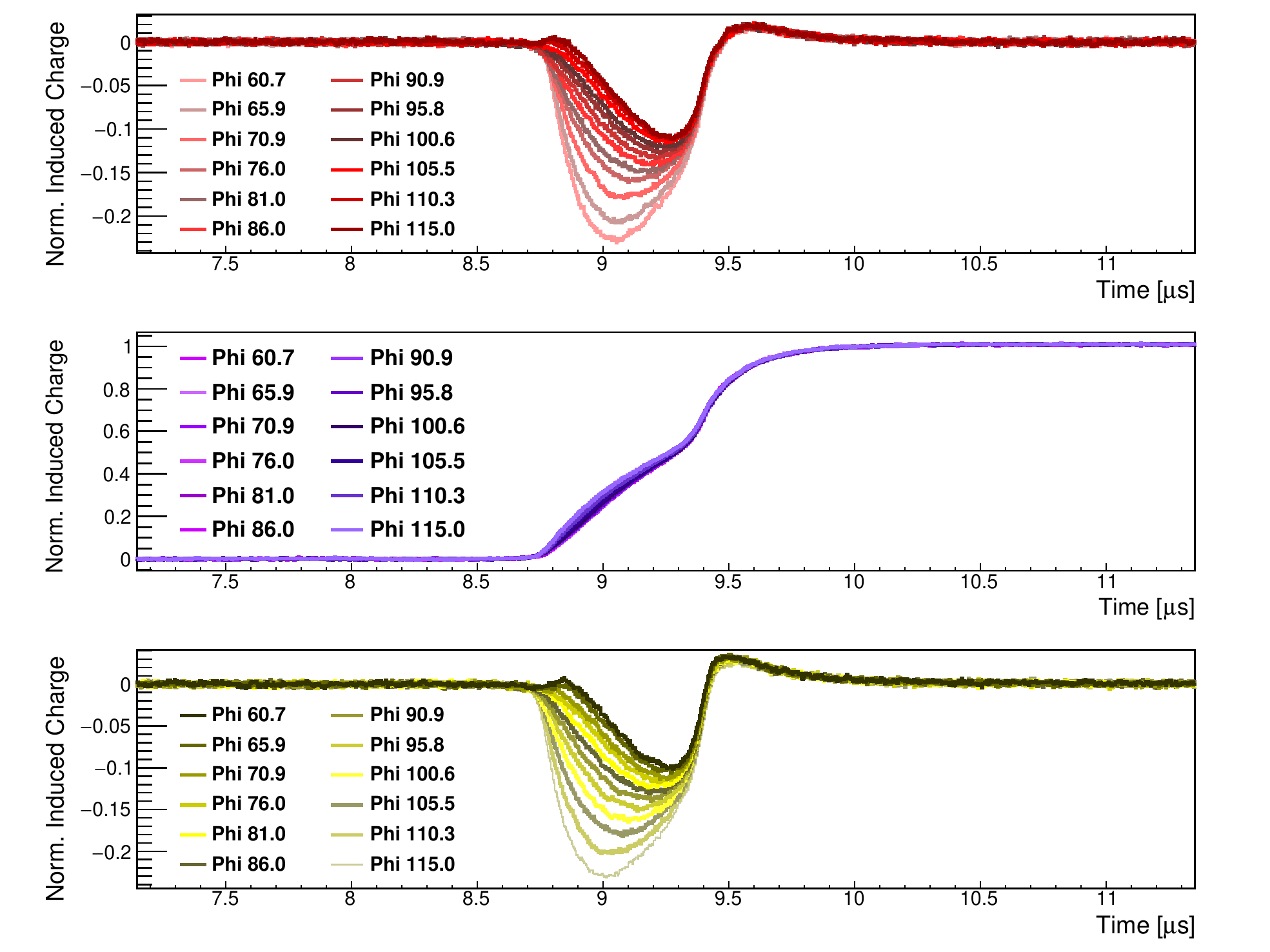}}
\vspace{-9cm} \hspace {9 cm} segment~1
\vskip 2.8cm  \hspace {9 cm} segment~4
\vskip 3.0cm  \hspace {9 cm} segment~2
\vspace{1.8cm}
\caption {From top to bottom:
          The super-pulses in segments~1,4,2 from the data set {\bf Ssc-B}
          in the range $60^o < \phi < 120^o$ for the 81\,keV line,
          adapted from~\cite{MTh:Schuster2017}.
          All pulses are normalised to an amplitude of 1 
          in the collecting segment~4. 
}
\label{fig:pos:pulses}
\end{figure}

\begin{figure}[h]
\centering
 \begin{subfigure}{.48\textwidth}
  \centering
  \includegraphics[width=1.0\textwidth]
           {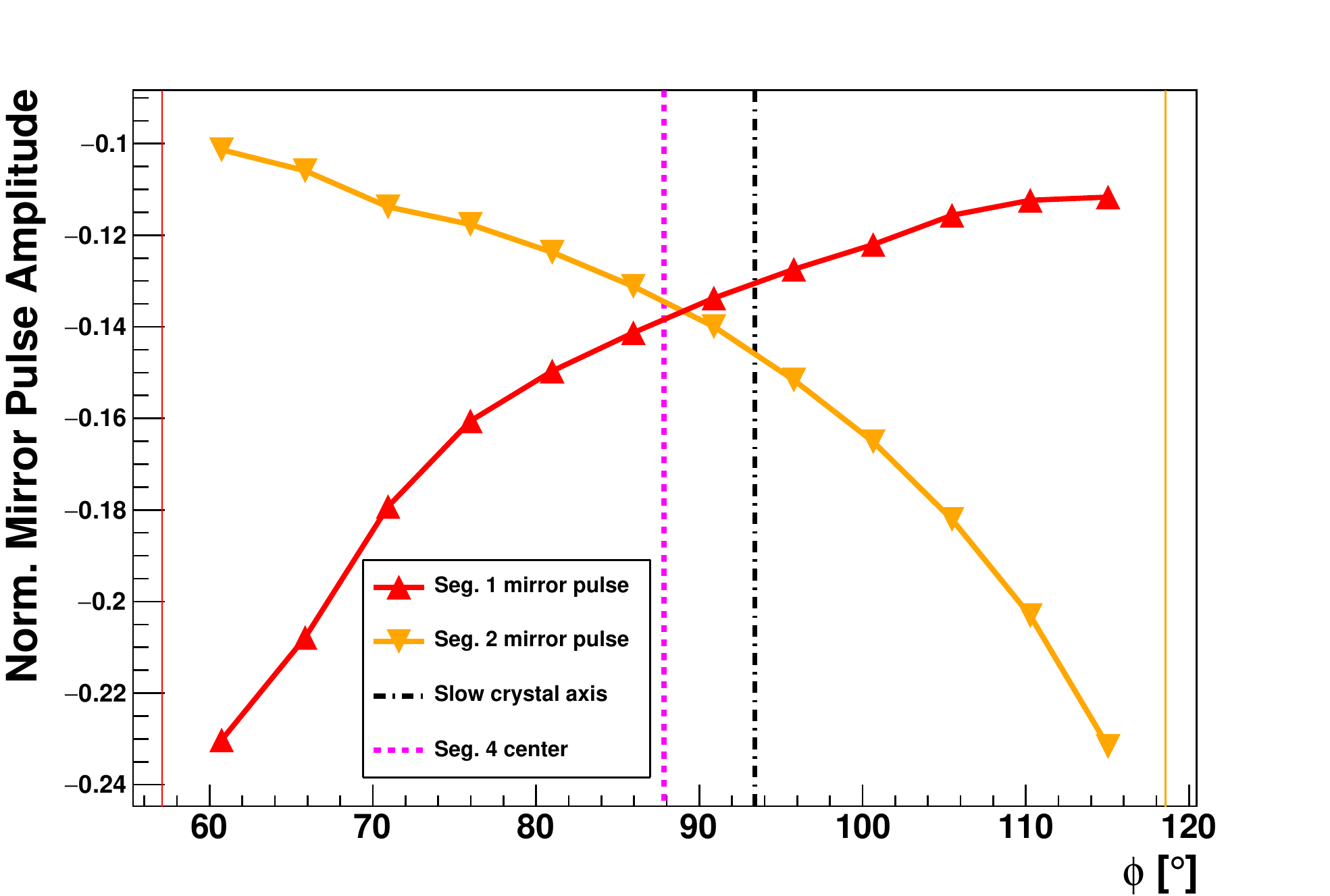}
  \end{subfigure}%
 \begin{subfigure}{.48\textwidth}
  \centering
  \includegraphics[width=1.0\textwidth]
            {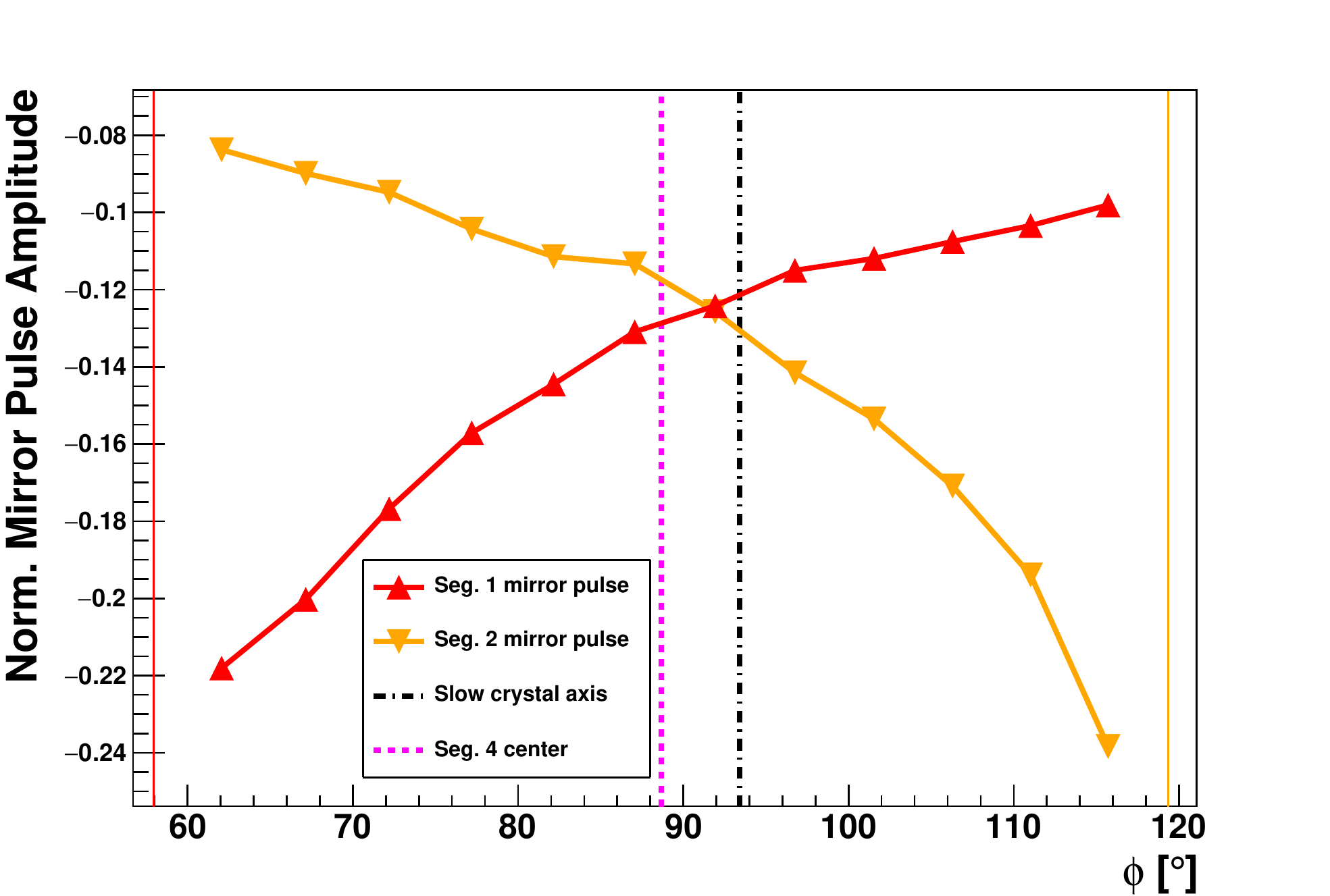}
  \end{subfigure}%
\vspace{0.5cm}
\caption{The mirror pulse amplitudes of the 81\,keV super-pulses 
         for the {\bf Ssc-B} (left) and {\bf Tsc} (right) scans 
         for the area of segment~4 between segment~1 
         and segment~2. Also indicated are the segment boundaries
         (solid vertical lines), the centre of segment~4 (dotted vertical line)
         and the location of the slow axis (dashed-dotted vertical line).  
         The data points are connected with straight lines to guide the
         eye.}
\label{fig:avg-mp}
\end{figure}

\begin{figure}[!h]
\centering
 \begin{subfigure}{.48\textwidth}
  \centering
  \includegraphics[width=1.0\textwidth]
           {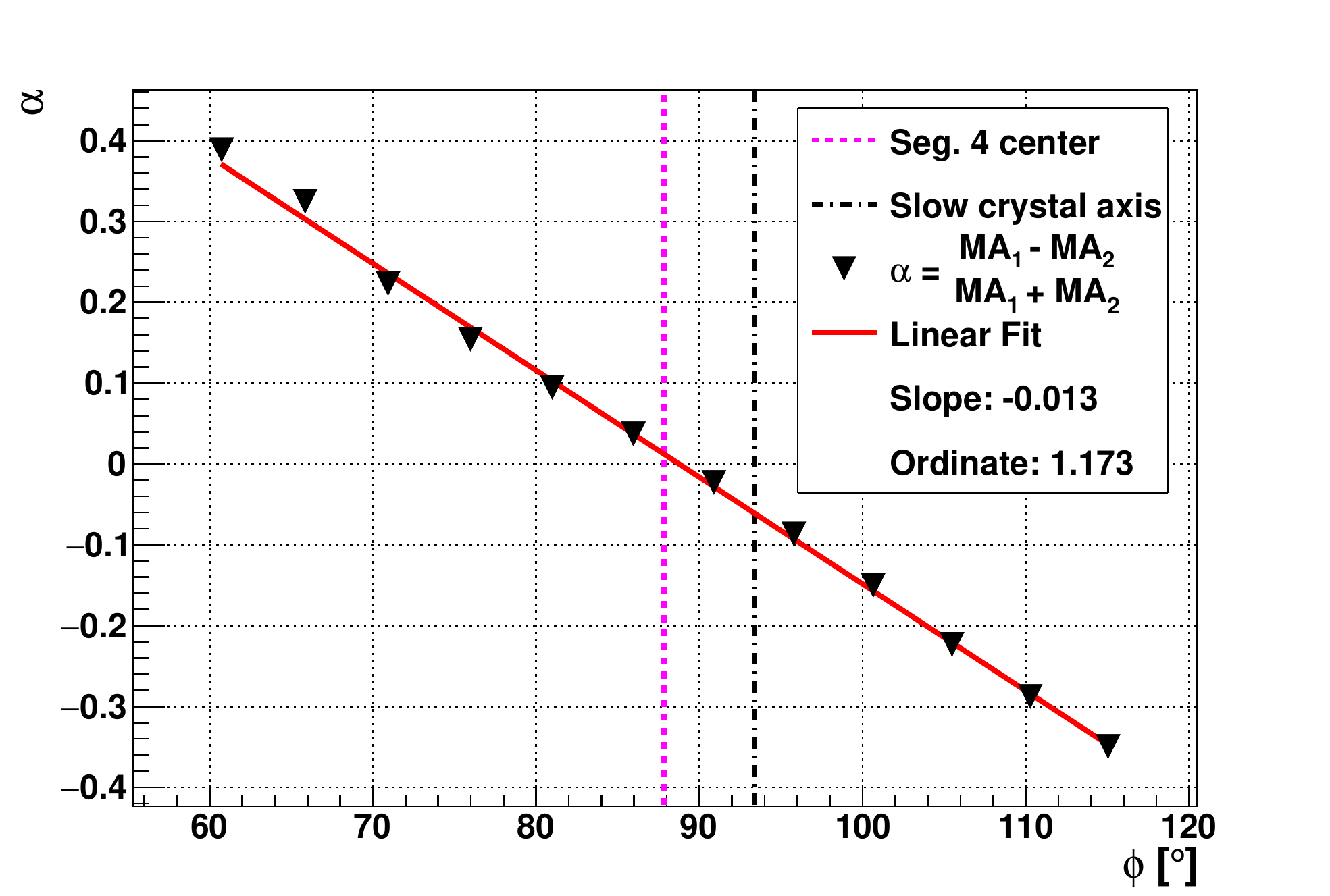}
  \end{subfigure}%
 \begin{subfigure}{.48\textwidth}
  \centering
  \includegraphics[width=1.0\textwidth]
            {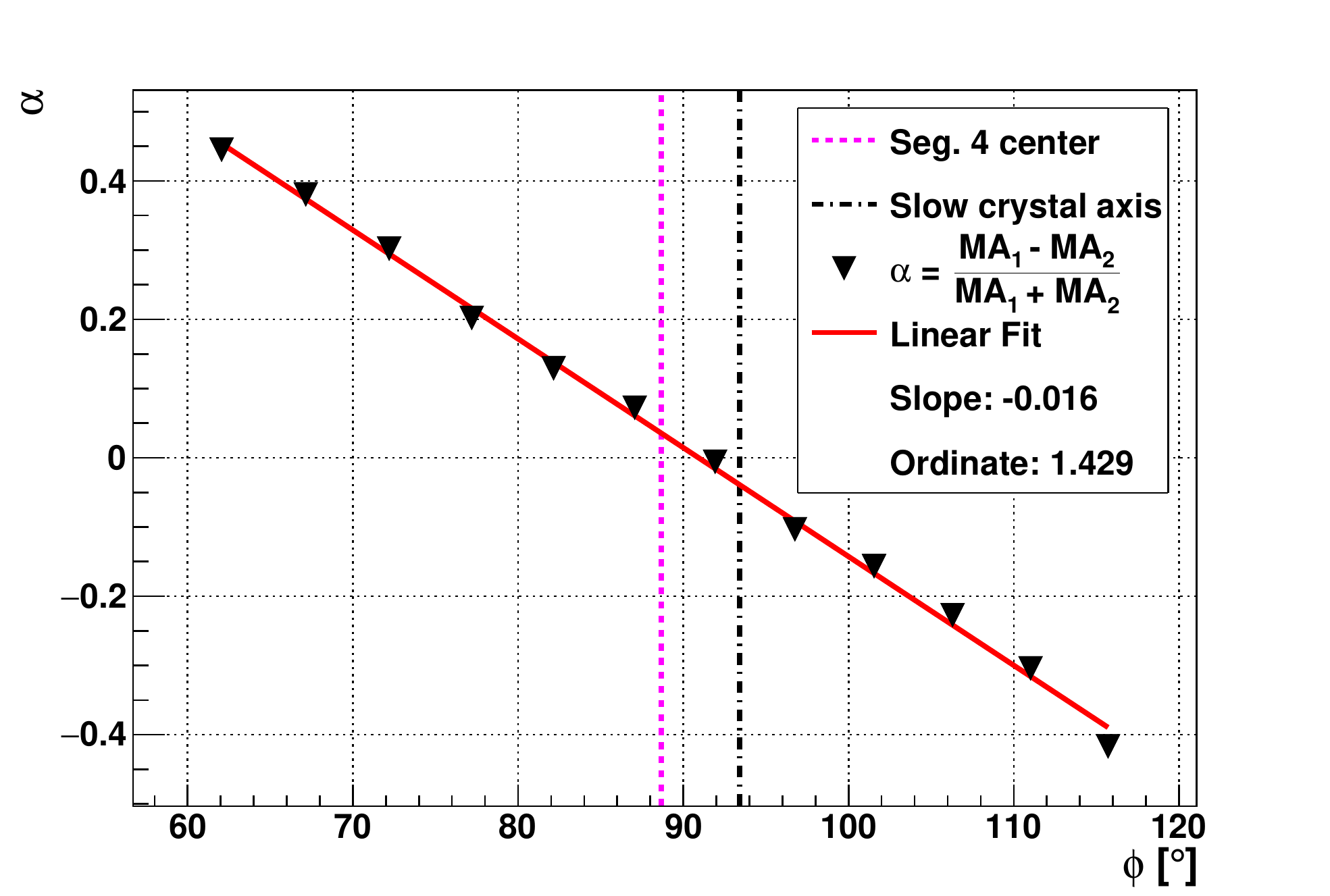}
  \end{subfigure}%
\vspace{0.5cm}
\caption{The asymmetries of mirror pulse amplitudes of the super-pulses 
         for the {\bf Ssc-B} (left) and {\bf Tsc} (right) scans for the area of 
         segment~4 between segment~1 
         and segment~2.  Also shown are linear fits to the data.
         }
\label{fig:asy-mp}
\end{figure}

The part of segment~4 between segments~1 and~2 was chosen to investigate
the phenomenon. The super-pulses from the 81\,keV line 
for $60^o < \phi < 120^o$ from 
the {\bf Ssc-B} data set are shown in Fig.~\ref{fig:pos:pulses} for 
the segments~1,4,2. The mirror pulses are all negative because only
the drift of the electrons is seen.
The MA values
of the pulses decrease for
segment~1 as the source moves away towards segment~2. At the same time
the MA values increase in segment~2. Both segments also show that
not only the amplitude of the mirror pulse changes but also the time at which
the amplitude is reached.
The pulse in the collecting segment~4 only changes moderately. This moderate
change is due to the
influence of the slow axis which is contained in that sector of segment~4.

The amplitudes, MA$_1$ and MA$_2$, of the pulses depicted in 
Fig.~\ref{fig:pos:pulses} are shown in the left panel of 
Fig.~\ref{fig:avg-mp}. 
The right panel depicts equivalent data from the
top-scan {\bf Tsc}.
Also indicated in the figure are the segment boundaries, the centre of
segment~4 and the location of the slow axis. Due to the influence of 
the slow axis,
the cross-over point between the two segment amplitudes is not at the
segment centre. Trajectories are bent
towards the slow axis and thus the cross-over point is pulled towards the
slow axis. The effect is larger for the {\bf Tsc} data because from 
$z=40$\,mm and $r=30$\,mm the inwards drift affected by the slow axis passes 
through a relatively low field a bit longer than for the {\bf Ssc-B} data at
$z=20$\,mm and $r=37.5$\,mm.

In order to reconstruct the position of the source, a simple
asymmetry, $\alpha$, was used:
\begin{equation}
\label{eq:asym}
 \alpha = \frac {MA_1 - MA_2}{MA_1 + MA_2} ~~.
\end{equation}
The asymmetries for the mirror pulse amplitudes 
together with linear fits 
are shown in Fig.~\ref{fig:asy-mp}. There was no attempt made
to provide statistical or $a~priori$ systematical uncertainties.
The linear fits are quite good. However, they provide 
different slopes and ordinates. The charge carrier trajectories
are very different for side and top-scans. Considering this, the
differences are expected.
In general, the trajectories are very dependent on the $z$-position 
for side-scans and $r$-position for top-scans. Thus, a simple asymmetry 
like $\alpha$ can only reconstruct the $\phi$ 
of surface events to about 10~degrees~\cite{MTh:Schuster2017} if there
is no information on $z$ ($r$) available.

\section{Charge Losses around the Core Contact}
\begin{figure}[h]
\vspace{-0.3cm} 
\centerline{
\includegraphics[width=0.8\textwidth]
   {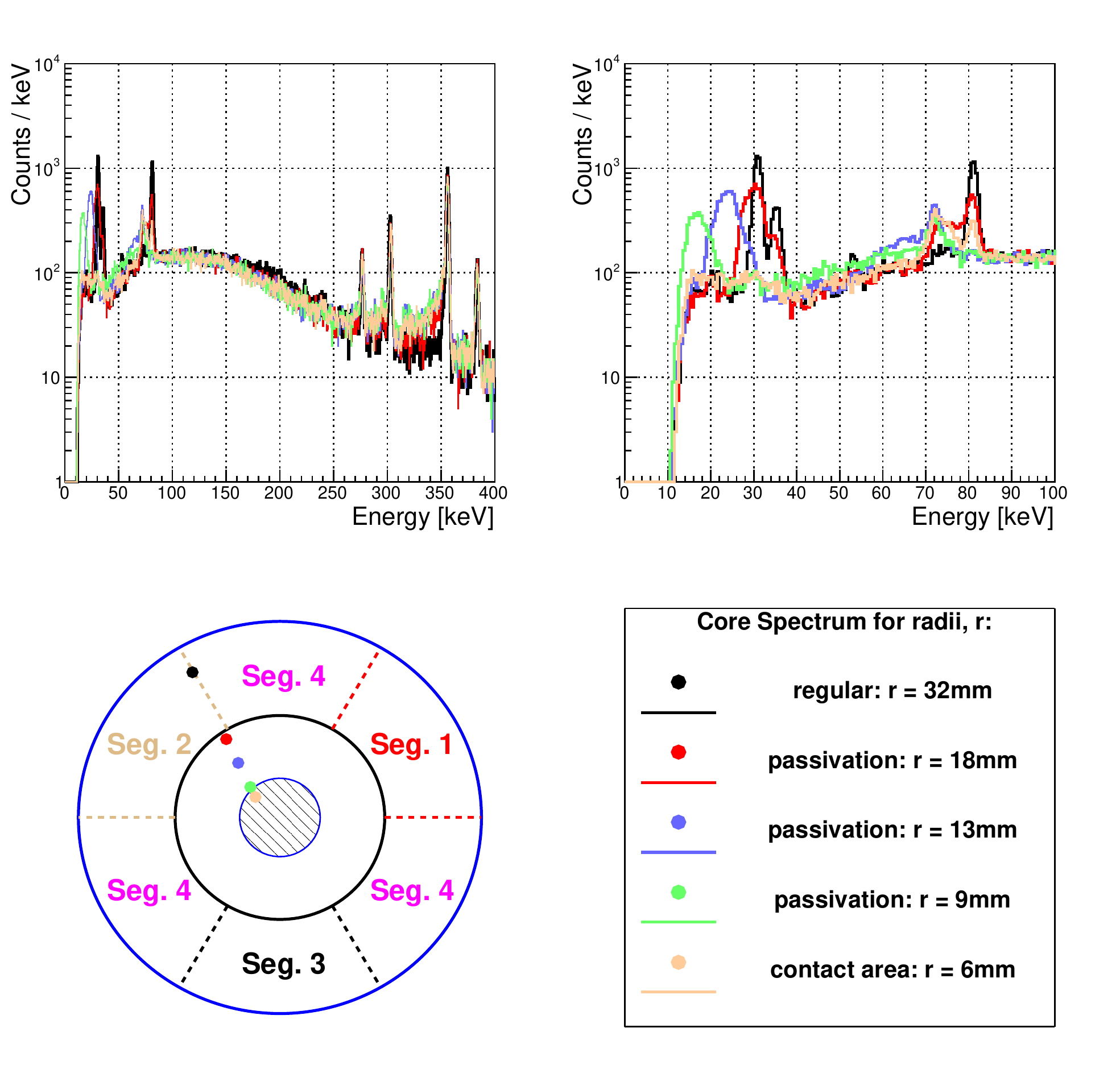}}
\vspace{0.5cm}
\caption {The core spectra for top-scan locations on the contact, 
          the passivation
          ring and the ``regular'' area of the detector, 
          adapted from~\cite{MTh:Schuster2017}.
          The spectra are shown from 10 to 400\,keV in the top left and
          from 10 to 100\,keV in the top right. The spectra are 
          coloured (shaded) 
          according to the legend. The scan locations are shown in the
          bottom left.  
}
\label{fig:core:losses}
\end{figure}

The core contact and the surrounding passivation ring
were investigated especially to look for possible charge losses. 
The core spectra for selected irradiation points at different radii
are shown in Fig.~\ref{fig:core:losses}.

The reference spectrum at $r=$\,32\,mm shows the expected peak at 81\,keV
and the double peak at 31\,keV and 35\,keV from the $^{133}$Ba source. 
In addition,
higher energy lines are shown. As soon as part of the beam spot reaches the
passivation ring, part of the events show a reduced energy. For the source
locations where the beam spot is fully contained on the passivation ring, 
all events show
a reduced energy. When part of the beam spot illuminates the core contact
some events are again observed at the nominal energy of 81\,keV.
The effect of the passivation ring is even stronger for the 31\,keV and 35\,keV
lines. These photons have even less penetration power than the 81\,keV
photons and, thus, it is not surprising that they are affected more.
It is noteworthy that these 31\,keV and 35\,keV peaks cannot 
be observed when the 
beam spot illuminates the core contact which is Lithium drifted.

The higher energy lines are also affected. They show pronounced low-energy 
shoulders which are caused by events with a shallow interaction point.
However, for most events the expected energy is recorded.

\begin{figure}[h]
\vspace{-0.3cm} 
\centerline{
\includegraphics[width=0.95\textwidth]
   {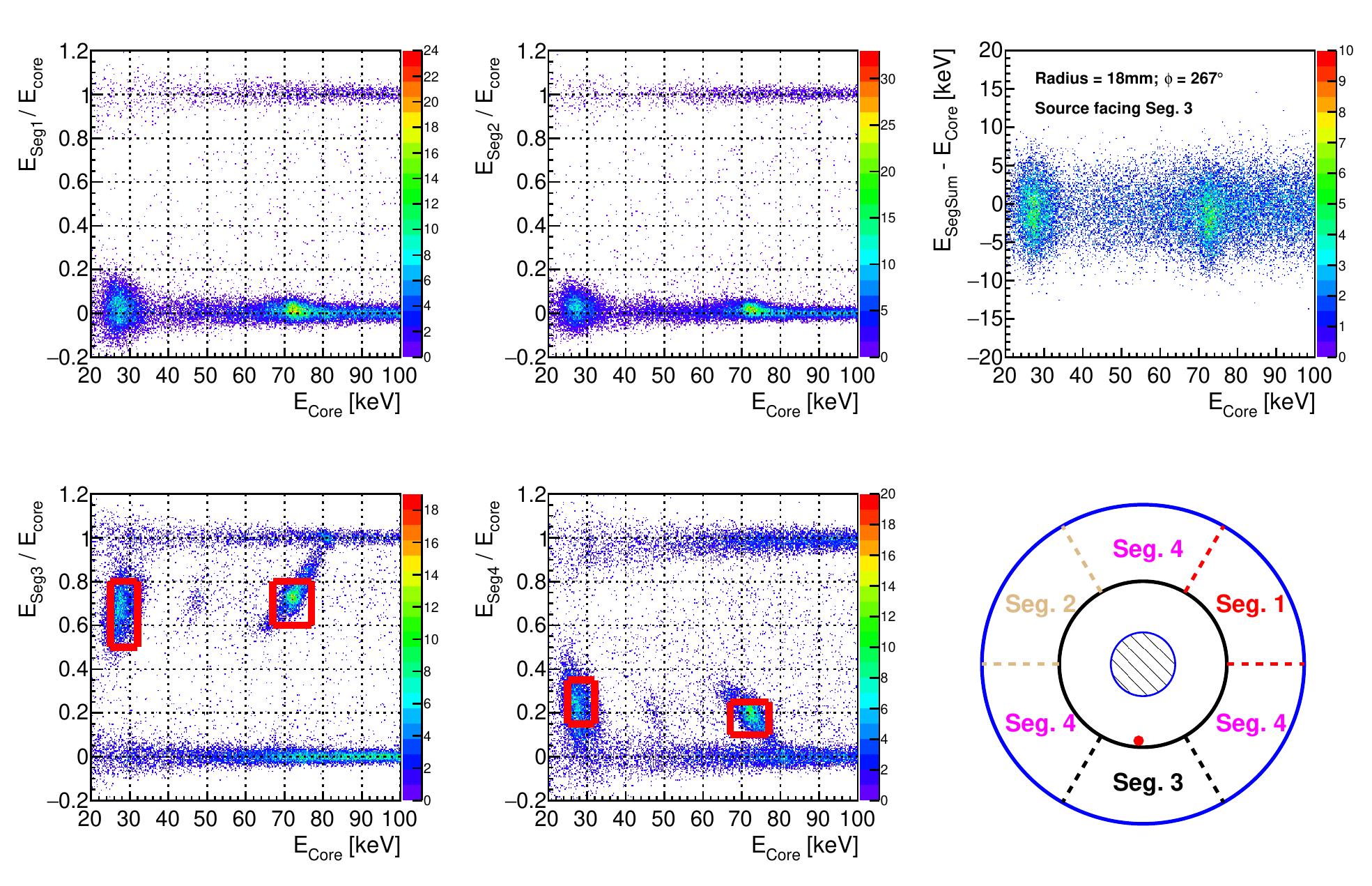}}
\vspace{0.5cm}
\caption {Distributions of the ratios of segment energies, $E_{\rm Seg\,i}$, 
          divided by the core energy $E_{\rm core}$ vs. $E_{\rm core}$ for 
          $i=1$ to~4 in the four left panels.
          The top right panel shows the difference between
          $\sum E_{\rm Seg\,i}$ and $E_{\rm core}$ vs. $E_{\rm core}$. 
          The graphic at the bottom right indicates the source position. 
         }   
\label{fig:segratio:losses}
\end{figure}

The behaviour as observed in Fig.~\ref{fig:core:losses} was the same for
a given $r$, independent of $\phi$.
Figure~\ref{fig:segratio:losses} provides information on
the segment energies for individual events
for a beam spot partially illuminating the outer edge of
the passivation ring. 
Even though gammas in the beam spot deposited their energy very close 
to segment~3, all segments show events where the charge is actually collected
in these segments. For these events, $E_{\rm Seg\,i}/E_{\rm core} \approx 1$.
Most of them can be associated with background.  
The relevant features are observed in segments~3 and~4. Part of the deposited
charge is not collected in segment~3, but in segment~4. 
This is clearly visible for the 81\,keV line and the effect is stronger
for the 31\,keV and 35\,keV lines.
This indicates that the drift paths right underneath the passivation ring
are irregular and that possibly a surface channel forms.
The charge loss observed in the core is also observed in the sum of segment
energies.
\begin{figure}[h]
\centerline{
\includegraphics[width=0.62\textwidth]
   {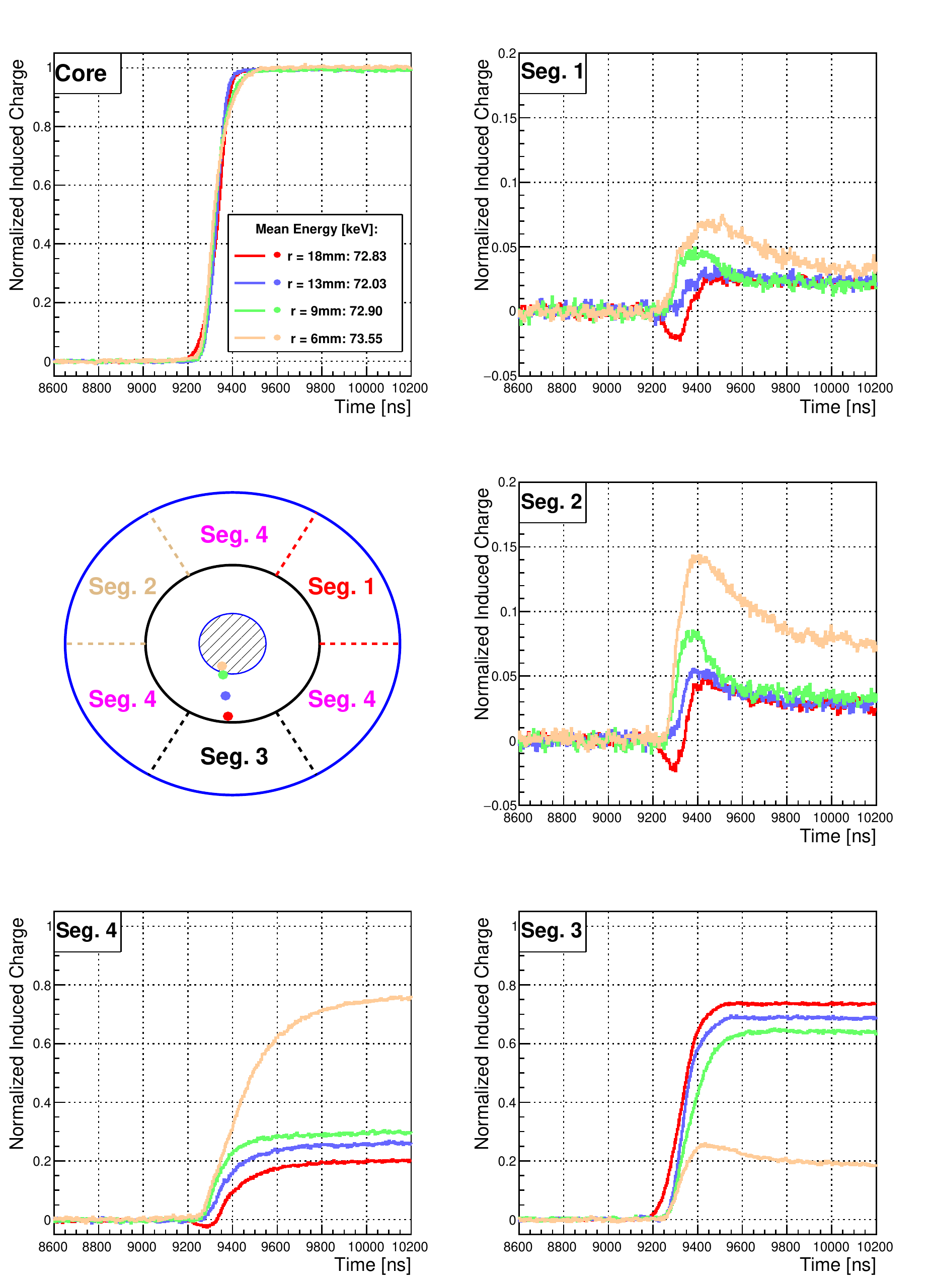}}
\caption {Super-pulses for 81\,keV events as indicated by the small
          boxes in Fig.~\ref{fig:segratio:losses}. The top left panel
          depicts core pulses, segments~1 to~3 are shown from top to
          bottom on the right, segment~4 pulses are depicted on the
          bottom left. 
          The graphic at the centre left indicates the source positions.
          Source positions and line appearance are matched 
          in colour (shading).   
}
\label{fig:pulse:losses}
\end{figure}

The events within the red boxes were identified as events with charge losses
and super-pulses of these events were formed.
Such super-pulses for the core and all segments for different 
radii are shown in Fig.~\ref{fig:pulse:losses}. 
They provide some insight on
what happens for interactions underneath the passivation layer.

For interaction points underneath the passivation ring or contact,
the holes are not immediately absorbed like underneath a segment
but have to drift a considerable distance.
As the interaction points move towards the core contact this hole drift
becomes longer and longer. As a result, charge is also collected
in segment~4 and the mirror pulses 
in segments~2 and~3 become positive.
The charge collected in segment~3 drops steadily while the charge collected
in segment~4 increases. Already at the centre of the ring, more energy is 
collected in segment~4 which is considerably larger and has a larger 
capacitance.

Even for the source position close to segment~3, 
the mirror pulses in 
segments~1 and~2 turn positive after the fast drift of the electrons 
towards the core contact has ended. This indicates that the holes drift
much slower in this region than the electrons. In addition, some holes
are pulled towards segment~4 and thus come closer to segments~1 and~2.
The core pulses are only
slightly affected because the weighting potential for the core drops
quickly for increasing $r$.

The positive mirror pulses in segments~1 and~2 do not return to 
the baseline. This indicates that at least part of the holes are trapped and do not reach
any segment electrode. This creates the charge loss.
When adding the mirror pulse amplitudes to the collected charges, the
line energy is roughly recovered.
These observations hint at the formation of a surface channel where holes
drift very slowly right underneath the passivation layer and have
a finite probability to get trapped.

\section{Summary and Outlook}

A four--fold $\phi$-segmented n-type BEGe detector was tested extensively at the
``Max-Planck-Institut f{\"u}r Physik'' in Munich. The segmentation did not
influence the detector performance as measured through the core contact.
The result of an A/E pulse-shape analysis was compatible with the results
obtained for standard BEGe detectors.
   
The location of detector segments and crystal axes could be 
reliably established through low-energy gamma surface scans
using super-pulses.
Mirror pulses in non-collecting segments were investigated and 
their super-pulses were used
to reconstruct the $\phi$ position of the interactions of 
low-energy gammas.
Mirror pulses were also used to investigate charge losses underneath the
core contact and the passivation ring. They can be attributed to 
low fields, surface channel effects and trapped holes.

After the series of measurements presented here, the detector was mounted
in a temperature controlled cryostat. The influence of the temperature
on the drift speed along different axes will be studied using detailed
pulse-shape information. 
The super-pulses from the collecting segment~4 as shown in 
Fig.~\ref{fig:pos:pulses} clearly reveal the different drift zones
in the detector. The electrons drift inwards first and
then reach the high field zone and are pulled upwards to the contact.
This allows the separation of drift times along different axes 
and, thus, a detailed study of the mobility tensor depending on
all three crystallographic axes.
In addition, the temperature dependence of the observed charge losses
will be investigated. 

Segmentation is an excellent tool to study detector properties.
The detector presented here is based on an n-type crystal. A similar
detector based on a p-type crystal is currently under construction.

\vfill

\newpage

{\bf Bibliography}

\bibliography{resources}

\begin{thebibliography}{10}
\expandafter\ifx\csname url\endcsname\relax
  \def\url#1{\texttt{#1}}\fi
\expandafter\ifx\csname urlprefix\endcsname\relax\def\urlprefix{URL }\fi
\expandafter\ifx\csname href\endcsname\relax
  \def\href#1#2{#2} \def\path#1{#1}\fi

\bibitem{Vetter2007:RecentDevelopmentsFabrication}
K.~Vetter, \href{https://doi.org/10.1146/annurev.nucl.56.080805.140525}{Recent
  developments in the fabrication and operation of germanium detectors}, Annual
  Review of Nuclear and Particle Science 57~(1) (2007) 363--404.
\newblock \href
  {http://arxiv.org/abs/https://doi.org/10.1146/annurev.nucl.56.080805.140525}
  {\path{arXiv:https://doi.org/10.1146/annurev.nucl.56.080805.140525}}, \href
  {http://dx.doi.org/10.1146/annurev.nucl.56.080805.140525}
  {\path{doi:10.1146/annurev.nucl.56.080805.140525}}.
\newline\urlprefix\url{https://doi.org/10.1146/annurev.nucl.56.080805.140525}

\bibitem{Akkoyun2011:AGATAAdvancedGamma}
S.~Akkoyun, et~al., {AGATA -- Advanced Gamma Tracking Array}, Nucl. Instr. and
  Meth. A 668 (2012) 26.
\newblock \href {http://arxiv.org/abs/1111.5731v2} {\path{arXiv:1111.5731v2}},
  \href {http://dx.doi.org/10.1016/j.nima.2011.11.081}
  {\path{doi:10.1016/j.nima.2011.11.081}}.

\bibitem{Abgrall2014:MajoranaDemonstratorNeutrinoless}
N.~Abgrall, et~al., {The Majorana Demonstrator Neutrinoless Double-Beta Decay
  Experiment}, Adv. High Energy Phys. 2014 (2014) 365432.
\newblock \href {http://arxiv.org/abs/1308.1633} {\path{arXiv:1308.1633}},
  \href {http://dx.doi.org/10.1155/2014/365432}
  {\path{doi:10.1155/2014/365432}}.

\bibitem{Ackermann2013:Gerdaexperimentsearch}
K.-H. Ackermann, et~al.,
  \href{https://doi.org/10.1140/epjc/s10052-013-2330-0}{{The GERDA experiment
  for the search of 0$\nu$$\beta$$\beta$ decay in 76Ge}}, Eur. Phys. J. C
  73~(3) (2013) 2330.
\newblock \href {http://dx.doi.org/10.1140/epjc/s10052-013-2330-0}
  {\path{doi:10.1140/epjc/s10052-013-2330-0}}.
\newline\urlprefix\url{https://doi.org/10.1140/epjc/s10052-013-2330-0}

\bibitem{Aalseth2013:CoGeNTSearchLow}
C.~Aalseth, et~al., {CoGeNT: A Search for Low-Mass Dark Matter using p-type
  Point Contact Germanium Detectors}, Phys. Rev. D88~(1) (2013) 012002.
\newblock \href {http://arxiv.org/abs/1208.5737} {\path{arXiv:1208.5737}},
  \href {http://dx.doi.org/10.1103/PhysRevD.88.012002}
  {\path{doi:10.1103/PhysRevD.88.012002}}.

\bibitem{Agostini2015:Productioncharacterizationoperation}
M.~Agostini, et~al.,
  \href{https://doi.org/10.1140/epjc/s10052-014-3253-0}{{Production,
  characterization and operation of 76 Ge enriched BEGe detectors in GERDA}},
  Eur. Phys. J. C 75~(2) (2015) 39.
\newblock \href {http://dx.doi.org/10.1140/epjc/s10052-014-3253-0}
  {\path{doi:10.1140/epjc/s10052-014-3253-0}}.
\newline\urlprefix\url{https://doi.org/10.1140/epjc/s10052-014-3253-0}

\bibitem{Mertens2015:Majoranaexperience}
S.~Mertens, et~al., {MAJORANA Collaboration's Experience with Germanium
  Detectors}, Journal of Physics: Conference Series 606 (2015) 012005.
\newblock \href {http://dx.doi.org/10.1088/1742-6596/606/1/012005}
  {\path{doi:10.1088/1742-6596/606/1/012005}}.

\bibitem{Abt2010:Pulseshapesimulation}
I.~Abt, et~al., \href{https://doi.org/10.1140/epjc/s10052-010-1364-9}{Pulse
  shape simulation for segmented true-coaxial {HPGe} detectors}, Eur. Phys. J.
  C 68~(3-4) (2010) 609.
\newblock \href {http://dx.doi.org/10.1140/epjc/s10052-010-1364-9}
  {\path{doi:10.1140/epjc/s10052-010-1364-9}}.
\newline\urlprefix\url{https://doi.org/10.1140/epjc/s10052-010-1364-9}

\bibitem{MTh:Schuster2017}
M.~Schuster,
  \href{https://wwwgedet.mpp.mpg.de/publication/Schuster_MasterThesis.pdf}{Characterization
  of a segmented n-type broad energy germanium detector}, Master's thesis,
  Technische Universit{\"a}t M{\"u}nchen (2017).
\newline\urlprefix\url{https://wwwgedet.mpp.mpg.de/publication/Schuster_MasterThesis.pdf}

\bibitem{Abt2007:Characterizationfirsttrue}
I.~Abt, et~al.,
  \href{http://www.sciencedirect.com/science/article/pii/S0168900207005566}{{Characterization
  of the first true coaxial 18-fold segmented n-type prototype HPGe detector
  for the GERDA project}}, Nucl. Inst. and Meth. A 577 (2007) 574.
\newblock \href
  {http://dx.doi.org/http://dx.doi.org/10.1016/j.nima.2007.03.035}
  {\path{doi:http://dx.doi.org/10.1016/j.nima.2007.03.035}}.
\newline\urlprefix\url{http://www.sciencedirect.com/science/article/pii/S0168900207005566}

\bibitem{Pixie4}
XIA, Manual: 4-Channel 75 MHz PXI Digital Spectrometer (2013).

\bibitem{def:2014}
Struck Innovative Systeme GMBH,
  \href{http://www.struck.de/sis3316-2014-03-20.pdf}{SIS3316 Family 16 Channel
  VME Digitizer} (2014).
\newline\urlprefix\url{http://www.struck.de/sis3316-2014-03-20.pdf}

\bibitem{MTh:Hauertmann2017}
L.~Hauertmann,
  \href{https://wwwgedet.mpp.mpg.de/publication/thesis_final_20170330_compressed.pdf}{{Influence
  of the Metallization on the Charge Collection Efficiency of Segmented
  Germanium Detectors}}, Master's thesis, Technische Universit{\"a}t
  M{\"u}nchen (2017).
\newline\urlprefix\url{https://wwwgedet.mpp.mpg.de/publication/thesis_final_20170330_compressed.pdf}

\bibitem{PhD:Liao2016}
H.-Y. Liao,
  \href{http://nbn-resolving.de/urn:nbn:de:bvb:19-195195}{{Development of pulse
  shape discrimination methods for BEGe detectors}}, Ph.D. thesis,
  Ludwig-Maximilians-Universit{\"a}t M{\"u}nchen (2016).
\newline\urlprefix\url{http://nbn-resolving.de/urn:nbn:de:bvb:19-195195}

\bibitem{Reik1962:crystalaxis}
H.~G. Reik, H.~Risken, Drift velocity and anisotropy of hot electrons in n
  germanium, Phys. Let. 126 (1962) 1737.
\newblock \href {http://dx.doi.org/doi.org/10.1103/PhysRev.126.1737}
  {\path{doi:doi.org/10.1103/PhysRev.126.1737}}.

\end{thebibliography}

\end{document}